\documentclass[aps,ams,prx,reprint, superscriptaddress]{revtex4-2}
\usepackage{amsmath, amsfonts}
\usepackage{graphicx}
\usepackage{amssymb}
\usepackage{hyperref}
\usepackage{booktabs}
\usepackage{xcolor}
\providecommand{\bmhead}[1]{\section*{#1}}
\begin{document}
\title{First-Principles Quantum-Spectral framework \\ for Elementary Vortex Pinning in superconductors}

\author{Haozhe Shi}
\thanks{These authors are co-first authors and contributed equally to this work.}
\affiliation{Department of Physics, Fudan University, Shanghai 200433, China}
\affiliation{Key Laboratory of Computational Physical Sciences (Ministry of Education), State Key Laboratory of Surface Physics,
Fudan University, Shanghai 200433, China}

\author{Yuncheng Xie}
\thanks{These authors are co-first authors and contributed equally to this work.}
\affiliation{Department of Physics, Fudan University, Shanghai 200433, China}
\affiliation{Key Laboratory of Computational Physical Sciences (Ministry of Education), State Key Laboratory of Surface Physics,
Fudan University, Shanghai 200433, China}

\author{Tong Zhang}
\affiliation{Department of Physics, State Key Laboratory of Surface Physics and Advanced Material Laboratory,
Fudan University, Shanghai 200438, China}
\affiliation{Hefei National Laboratory, Hefei 230088, China}

\author{Weibin Chu}
\affiliation{Department of Physics, Fudan University, Shanghai 200433, China}
\affiliation{Key Laboratory of Computational Physical Sciences (Ministry of Education), State Key Laboratory of Surface Physics,
Fudan University, Shanghai 200433, China}

\author{Xin-Gao Gong}
\affiliation{Department of Physics, Fudan University, Shanghai 200433, China}
\affiliation{Key Laboratory of Computational Physical Sciences (Ministry of Education), State Key Laboratory of Surface Physics,
Fudan University, Shanghai 200433, China}

\begin{abstract}
The critical current of a type-II superconductor is controlled by vortex pinning, whose microscopic input is the elementary pinning force. Scanning tunneling spectroscopy has shown that a defect pins a vortex by reorganizing the Caroli--de Gennes--Matricon (CdGM) states in its core, but why this spectral reorganization amounts to a pinning force has lacked a quantum-mechanical, first-principles account. Here we establish a transferable first-principles computational framework for elementary vortex pinning, in which defect-resolved DFT/Wannier electronic structures are embedded into a finite-box projected Bogoliubov--de Gennes free-energy formalism to convert quasiparticle spectral reorganization into vortex-pinning energies and forces. Using this framework, we confirm that the defect-induced reorganization of the vortex-core spectrum is the microscopic origin of the elementary pinning force. The force is evaluated as a finite-box vortex-insertion free energy whose four-configuration subtraction isolates the meV-scale interaction from much larger backgrounds. With the superconducting gap scale and vortex-core profile fixed from experiments, the FeSe Fe-site vacancy reproduces the microscopic STM force scale together with the measured spectral reorganization. All five point defects in FeSe and FeTe pin attractively, with FeTe Fe-site vacancy strongest. Elementary vortex pinning thereby becomes a computable electronic-structure quantity, opening the first-principles screening of point defects toward higher critical currents.
\end{abstract}

\maketitle

\section{Introduction}

Type-II superconductors derive their technological value from their ability to carry dissipationless currents in a magnetic field~\cite{Larbalestier2001HighTc}. That ability, however, is never guaranteed by superconductivity alone. Once magnetic flux penetrates the material as quantized vortices~\cite{Abrikosov1957TypeII}, an applied current exerts a Lorentz force that drives vortex motion and generates energy loss. The practical current limit is therefore set not by pairing strength or transition temperature, but by how effectively defects immobilize vortices. In this sense, flux pinning is a critical bottleneck constraining the transition of superconductivity from a quantum state to a high-current technology suitable for practical application.

Despite decades of progress in vortex physics, the microscopic theory of flux pinning remains incomplete. Most materials strategies still rely on empirical defect engineering, phenomenological models, or mesoscale simulations with fitted inputs~\cite{Bean1964CriticalState,AndersonKim1964FluxCreep,DewHughes1974FluxPinning,LarkinOvchinnikov1979CollectivePinning,Blatter1994Vortices}. What is missing are the micro-scale descriptive parameters that enable predictive design:
namely, the elementary pinning energy and the corresponding elementary pinning force associated with specific defects~\cite{Thuneberg1984ElementaryPinning,Hyun1987ElementaryForce}. These quantities measure the binding of an individual vortex to an individual defect and provide the fundamental connection between atomic-scale disorder and the macroscopic critical current. In the absence of a reliable first-principles route to these quantities, the identification of effective pinning centers has remained largely empirical.

This gap has become increasingly striking in light of recent scanning tunneling spectroscopy experiments, which resolve individual vortex cores and the Caroli--de Gennes--Matricon (CdGM) states bound within them~\cite{Caroli1964CdGM,Hess1989STMFluxLattice,Chen2020FeSeCdGM,Zhang2021LiFeOHFeSeVortexModes} and directly visualized how a point defect pins a vortex by reorganizing that core spectrum~\cite{Zhang2023LiFeOHFeSeImpurity,Chen2024PRXPinning}. These observations indicate that vortex pinning is not merely a coarse consequence of local gap suppression or elastic vortex-line energetics, but an intrinsically quantum electronic process encoded in the defect-modified quasiparticle spectrum. Yet a central question has remained open: why does this spectral reorganization amount to a measurable pinning force, and can that force be predicted quantitatively from the electronic structure of a real material? Answering it is essential both for interpreting the spectroscopy and for turning microscopic insight into a design principle.

A first-principles account of elementary vortex pinning has been difficult for two reasons. First, the relevant interaction energy is extremely small, of order meV, while it is embedded in much larger electronic free-energy backgrounds, so a direct calculation cannot cleanly isolate the pinning signal. Second, vortex physics belongs to the thermodynamic limit, whereas any realistic electronic-structure calculation is performed in a finite computational cell, and the finite-size background can easily overwhelm the subtle vortex--defect interaction.

Here we overcome these barriers with a first-principles framework that computes the vortex--defect binding energy and the elementary pinning force directly from the electronic structure. Once the superconducting gap scale and vortex-core profile are fixed from experiment, the force is obtained without additional empirical fitting of the vortex--defect interaction. The approach combines a realistic description of the defect electronic structure with a projected Bogoliubov--de Gennes (BdG) free-energy formalism~\cite{deGennes1966Superconductivity,GygiSchluter1991VortexBdG} designed to isolate the vortex--defect interaction while canceling the dominant finite-size background with high precision. This makes it possible to translate the spectroscopic reorganization of vortex-core states into a quantitative force, and to do so for specific point defects in specific superconductors.

We apply this framework to point defects in FeSe and FeTe, two layered iron-chalcogenide superconductors whose superconductivity is realized in the two-dimensional limit~\cite{Kamihara2008IronBased,Huang2017LiFeOHFeSeFilm}. For FeSe, where defect-resolved vortex spectroscopy provides a stringent benchmark, the calculation reproduces the observed reconstruction of core states and yields an elementary pinning force of \(1.18\times10^{-4}\,\mathrm{N\,m^{-1}}\) for an Fe-site vacancy, on the same scale as the scanning tunneling value~\cite{Chen2024PRXPinning}. We then extend the same calculation to FeTe, recently reported to superconduct but still under debate~\cite{Qin2021FeTeBi2Te3,Ren2021OxygenFeTe,Yan2026StoichiometricFeTe}, as a predictive target. Across all five point defects considered the pinning is attractive, with the Te vacancy in FeTe the strongest. Comparing the spectral contribution with that of order-parameter suppression shows that the dominant source of pinning is the defect-induced reorganization of the quasiparticle spectrum, while gap suppression changes the force only within the same order of magnitude.

\section{Finite-box pinning landscape}

Our first step is to make precise what the elementary pinning interaction is as a quantity that a finite first-principles calculation can evaluate, namely the energy and corresponding force with which a specified point defect acts on a vortex~\cite{Thuneberg1984ElementaryPinning,Hyun1987ElementaryForce}. In the thermodynamic limit, this interaction can be defined conceptually by comparing two configurations: a vortex bound to the defect and the same vortex far from the defect.

However, as stressed in the Introduction, a direct comparison of these total free energies cannot resolve the meV-scale interaction in a finite box, where each
value is dominated by the defect background, the vortex self-energy, and
finite-size effect. We therefore compare vortex-insertion costs
instead: for each separation, the cost of inserting a vortex is evaluated
against a no-vortex reference built from the same Hamiltonian, and
these costs are compared between pinned and reference configurations.

\begin{figure}[htbp]
\centering
\includegraphics[width=1.0\linewidth]{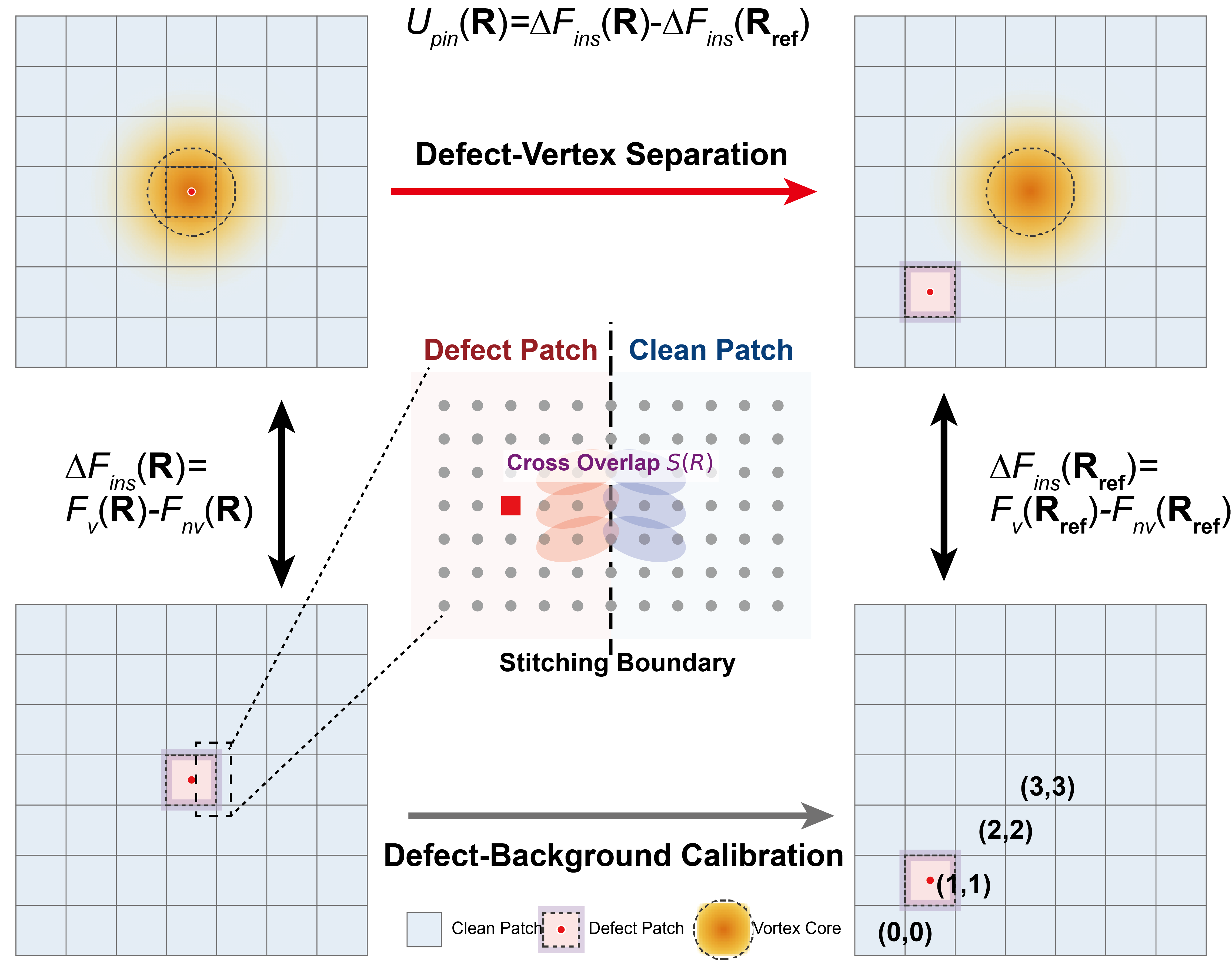}
\caption{Schematic of the finite-box vortex-insertion framework. The defect patch is separated from the vortex core by $\mathbf{R}$. The insertion cost $\Delta F_{\rm ins}(\mathbf{R})$ is the difference between vortex and no-vortex free energies. The pinning landscape \(U_{\rm pin}(R)\) is referenced to the same
defect patch placed at a buffered separation \(R_{\rm ref}\), away
from both the vortex core and the boundary.}
\label{fig:method_schematic}
\end{figure}

Let \(\mathbf R\) be the in-plane separation between a fixed vortex center and an embedded point-defect patch(see Fig.~\ref{fig:method_schematic}).  For the same defect Hamiltonian, we compute the projected BdG free energy \(F_{\rm v}(\mathbf R)\) with a vortex order-parameter texture and \(F_{\rm nv}(\mathbf R)\) for the corresponding no-vortex superconducting reference.  The vortex-insertion cost is defined as
\begin{equation}
\Delta F_{\rm ins}(\mathbf R)=F_{\rm v}(\mathbf R)-F_{\rm nv}(\mathbf R).
\label{eq:vortex_insertion_cost}
\end{equation}
Because the two terms are evaluated with the same basis, filling, cutoff, and BdG free-energy convention, this subtraction removes the defect background unrelated to vortex insertion.

The finite-box pinning landscape is then obtained by comparing the vortex-insertion cost at different defect--vortex separations:
\begin{equation}
\begin{aligned}
U_{\rm pin}(\mathbf R)
&=
\Delta F_{\rm ins}(\mathbf R)-\Delta F_{\rm ins}(\mathbf R_{\rm ref})  \\
&=
\left[F_{\rm v}(\mathbf R)-F_{\rm nv}(\mathbf R)\right]
-
\left[F_{\rm v}(\mathbf R_{\rm ref})-F_{\rm nv}(\mathbf R_{\rm ref})\right].
\end{aligned}
\label{eq:pinning_landscape}
\end{equation}
Here \(\mathbf R_{\rm ref}\) is a buffered reference position inside the same finite box, chosen away from both the vortex core and the boundary.  With this convention \(U_{\rm pin}(\mathbf R_{\rm ref})=0\), and \(U_{\rm pin}(\mathbf 0)<0\) denotes attractive pinning. This four-configuration subtraction cancels the leading defect background and common vortex contribution, so that the remaining signal is dominated by the vortex--defect interaction.

The elementary force per unit vortex length is obtained from the local slope of the pinning landscape.  In the continuum limit one would write
\begin{equation}
\mathbf{f}_{p}(\mathbf R)
=
-\frac{1}{\ell_z}\nabla_{\mathbf R} U_{\rm pin}(\mathbf R),
\label{eq:continuous_pinning_force}
\end{equation}
where \(\ell_z\) is the physical thickness associated with the vortex segment.  In the present discrete finite-box calculations, we use the near-core finite difference
\begin{equation}
f_p^{\rm loc}
=
\frac{1}{\ell_z}
\left|
\frac{U_{\rm pin}(\mathbf R_3)-U_{\rm pin}(\mathbf R_2)}
{|\mathbf R_3-\mathbf R_2|}
\right| ,
\label{eq:local_pinning_force}
\end{equation}
where \(\mathbf R_3\) is the defect-centered configuration and \(\mathbf R_2\) is the nearest off-center configuration along the scan direction.  The quantity \(f_p^{\rm loc}\) is the elementary pinning force per unit vortex length.  It is the microscopic output passed to macroscopic pinning theories, not by itself a direct prediction of the sample-dependent \(J_c(B,T)\).

This four-configuration construction turns the elementary pinning interaction into a single, physically transparent quantity, the change in the cost of inserting a vortex when the defect is brought onto the core.  What remains is to specify the projected Bogoliubov--de Gennes (BdG) free energy \(F_x\) that enters it.  For each configuration \(x={\rm v},{\rm nv}\) we evaluate
\begin{equation}
F_x(\mathbf R;E_c)
=
-\frac12
{\rm Tr}
\left|
\mathcal H_{\rm BdG}^{(x)}(\mathbf R;E_c)
\right|
+
\frac{1}{g(E_c)}
\left\|
\Delta_{\mathcal C}^{(x)}(\mathbf R)
\right\|_F^2 .
\label{eq:bdg_free_energy}
\end{equation}
a quasiparticle spectral trace over a projected, cutoff-limited BdG subspace plus a cutoff-consistent mean-field counterterm, whose coupling \(g(E_c)\) is recalibrated at the same cutoff so that the two terms live in the same finite window and only their sum is physical.

The superconducting state enters as a calibrated BdG background carrying an imposed vortex texture, while the normal-state content of \(\mathcal H_{\rm BdG}\) follows from a first-principles pipeline in which clean and defective supercells are computed in density functional theory, wannierized into local Hamiltonians, and stitched into the finite box with a PAW-consistent overlap metric, a projected filling, and a projected pairing.  We defer this construction to the Appendix~\ref{app:implementation}; the physical content of Eq.~\eqref{eq:bdg_free_energy} itself is the subject of the next section.

\section{BdG free energy and the Ginzburg--Landau limit}

Equation~\eqref{eq:bdg_free_energy} is the central
thermodynamic object of the present calculation, and it is worth
understanding what physics it already contains. In the clean
uniform limit it reduces to the usual BCS condensation energy, and
allowing the order parameter to vary slowly yields the
Ginzburg--Landau (GL) condensation-energy functional through a
further long-wavelength expansion.  Away from that limit,
however, Eq.~\eqref{eq:bdg_free_energy} keeps the full
quasiparticle spectrum on which elementary pinning depends.

To see this explicitly, consider the clean, spatially uniform
limit of Eq.~\eqref{eq:bdg_free_energy}.  In this limit the
projected pairing matrix of Eq.~\eqref{eq:projected_pairing}
reduces to a uniform amplitude \(\Delta\), and the BdG
Hamiltonian decomposes into independent momentum blocks with
eigenvalues \(\pm E_k\),
\begin{equation}
    E_k=\sqrt{\xi_k^2+\Delta^2},
\end{equation}
with \(\xi_k\) the normal-state energy measured from the chemical potential. The quasiparticle trace then gives
\begin{equation}
    F_{\rm qp}(E_c)
    =
    -\frac12{\rm Tr}\,|\mathcal H_{\rm BdG}|
    =
    -\sum_{k\in\mathcal C}E_k ,
    \label{eq:uniform_qp_trace}
\end{equation}
where \(\mathcal C\) denotes the same symmetric cutoff window
used in the projected calculation.  Measuring the
superconducting free energy relative to the normal state,
\(E_k\to|\xi_k|\), gives
\begin{equation}
    F(E_c)
    =
    -\sum_{k\in\mathcal C}
    \left(E_k-|\xi_k|\right)
    +
    \frac{\mathcal N \Delta^2}{g(E_c)} .
    \label{eq:uniform_deltaF}
\end{equation}
Here \(\mathcal N\) is defined by the Frobenius norm of
the uniform projected pairing matrix,
    $\|\Delta_C\|_F^2
    \rightarrow
    \mathcal N \Delta^2$ ,
which keeps the notation compatible with the finite-dimensional
projected form of Eq.~\eqref{eq:bdg_free_energy}.

The cutoff-dependent coupling is fixed in the same projected
window by the stationarity of the uniform free energy,
\(\partial F/\partial\Delta=0\).  For \(\Delta\neq0\),
this gives
\begin{equation}
    \frac{\mathcal N}{g(E_c)}
    =
    \sum_{k\in\mathcal C}\frac{1}{2E_k}.
    \label{eq:uniform_gap_equation}
\end{equation}
In the present work this equation should be understood as the
cutoff calibration of the effective pairing strength.  The vortex and defect
textures used below are constrained BdG backgrounds, not fully
self-consistent extrema of the microscopic pairing functional.
The same calibrated \(g(E_c)\) is then carried over unchanged
to these nonuniform backgrounds.  For them the simple uniform
cancellation derived below no longer has a closed analytic form,
so cutoff stability must be checked numerically; this is
confirmed for the pinning energy in
Fig.~\ref{fig:cutoff_convergence}(c,d).

\begin{figure}[htbp]
\centering
\includegraphics[width=\columnwidth]{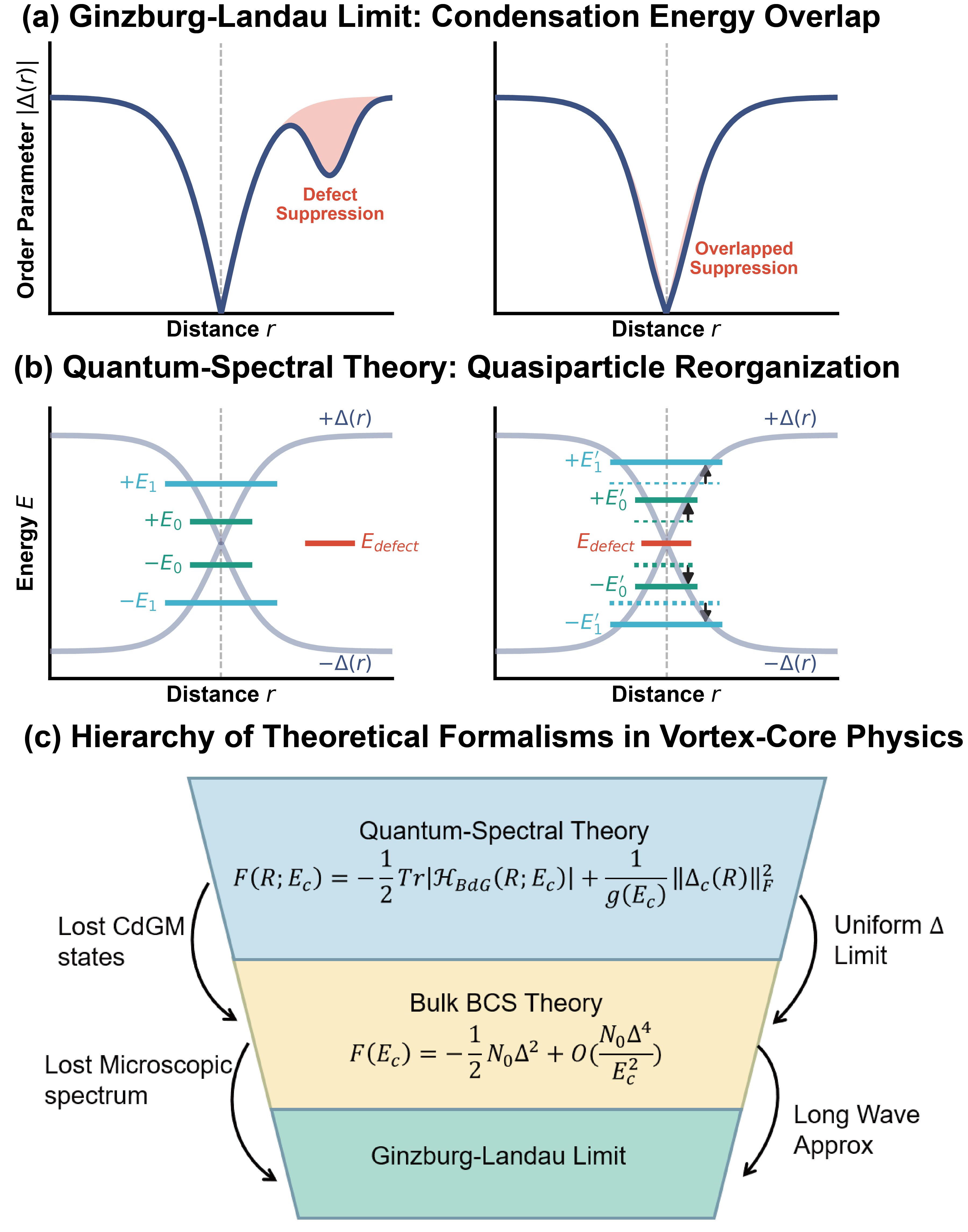} 
\caption{Comparison of macroscopic and microscopic vortex pinning mechanisms and their theoretical hierarchy. (a) In the macroscopic Ginzburg--Landau limit, vortex pinning is phenomenologically viewed as the spatial overlap of the order parameter suppression. (b) In our quantum-spectral theory, the pinning force originates microscopically from the hybridization and level repulsion of discrete CdGM bound states. (c) The theoretical reduction framework, illustrating how the full quantum-spectral description reduces to the bulk BCS and macroscopic GL limits via successive uniform-gap and long-wavelength approximations, which integrate out the spatial inhomogeneity and quasiparticle degrees of freedom.}
\label{fig:theory_hierarchy}
\end{figure}

Replacing the sum by an integral over a symmetric energy window
\(|\xi|\le E_c\), with \(N_0\) the single-spin density of states
of the projected normal bands at the Fermi level, gives for
\(E_c\gg\Delta\)
\begin{align}
    \sum_{k\in\mathcal C}
    \left(E_k-|\xi_k|\right)
    &=
    N_0\Delta^2
    \ln\frac{2E_c}{\Delta}
    +
    \frac12 N_0\Delta^2
    +
    O\!\left(\frac{N_0\Delta^4}{E_c^2}\right),
    \label{eq:uniform_qp_asymp}
    \\
    \frac{\mathcal N \Delta^2}{g(E_c)}
    &=
    N_0\Delta^2
    \ln\frac{2E_c}{\Delta}
    +
    O\!\left(\frac{N_0\Delta^4}{E_c^2}\right).
    \label{eq:uniform_ct_asymp}
\end{align}
Substituting Eqs.~\eqref{eq:uniform_qp_asymp} and
\eqref{eq:uniform_ct_asymp} into
Eq.~\eqref{eq:uniform_deltaF}, the cutoff-dependent logarithms
cancel between the quasiparticle trace and the counterterm.  The
remaining finite contribution is
\begin{equation}
    F(E_c)
    =
    -\frac12 N_0\Delta^2
    +
    O\!\left(\frac{N_0\Delta^4}{E_c^2}\right).
    \label{eq:bcs_condensation_energy}
\end{equation}
This is the zero-temperature BCS condensation energy in the
same normalization convention.  The important point is that the
negative condensation energy is not equal to the counterterm
alone.  It is the finite remainder after the quasiparticle
spectral trace and the pairing counterterm are combined with the
same cutoff.

The GL condensation-energy functional is obtained by taking a
further low-energy and long-wavelength limit of this same
fermionic free energy~\cite{Gorkov1959GL}.  Near \(T_c\), and
for an order parameter that varies slowly compared with
microscopic electronic length scales, integrating out the
quasiparticles and expanding the result in powers and gradients
of \(\Delta({\bf r})\) gives
\begin{equation}
\begin{aligned}
    F_{\rm GL}
    =
    \int d^3r\,\Big[
        &\alpha({\bf r})|\Delta({\bf r})|^2
        +
        \frac{\beta({\bf r})}{2}|\Delta({\bf r})|^4
        \\
        &+
        K({\bf r})|\mathbf D\Delta({\bf r})|^2
        +\cdots
    \Big].
\end{aligned}
    \label{eq:gl_functional}
\end{equation}
Thus the conventional GL condensation-energy picture is not a
separate assumption added to the BdG formalism.  It is the
smooth-spectrum, coarse-grained limit of the same microscopic
free-energy expression.

This observation defines precisely in what sense the present calculation goes beyond GL. As summarized in Fig.~\ref{fig:theory_hierarchy}, it is not beyond BdG mean-field theory, but rather goes beyond the coarse-grained GL condensation-energy approximation. In the conventional GL limit, pinning is phenomenologically treated as an overlap of condensation-energy suppression( Fig.~\ref{fig:theory_hierarchy}(a)). In contrast, our quantum-spectral theory resolves the underlying microscopic mechanism: the defect-induced hybridization and level repulsion of discrete CdGM bound states (Fig.~\ref{fig:theory_hierarchy}(b)). 

By evaluating Eq.~\eqref{eq:bdg_free_energy} before the quasiparticle spectrum is smoothed into local GL coefficients, we avoid the successive theoretical reductions (Fig.~\ref{fig:theory_hierarchy}(c)) that integrate out the crucial spatial inhomogeneity and quasiparticle degrees of freedom. A point defect perturbs the electronic structure on the atomic scale, and the vortex core contains discrete CdGM quasiparticle states. In a local GL core-overlap estimate these features are represented, at most, through coarse-grained coefficients rather than through explicit quasiparticle levels, whereas they are retained in the spectral trace of Eq.~\eqref{eq:bdg_free_energy}.

The same cancellation also explains the numerical behavior of
the cutoff dependence.  The quasiparticle trace and the
counterterm each depend on the projected window \(E_c\), but
their sum yields cutoff-stable free-energy differences, in
particular the vortex-insertion and pinning energies used here.
Therefore the separate quantities reported as
\(U_{\rm pin}^{\rm qp}\) and \(U_{\rm pin}^{\rm ct}\) should be
read only as diagnostic components of one regularized BdG
free-energy difference.  The physical quantities are the total
pinning energy and its spatial derivative, while the vortex-core
DOS shift discussed below provides the spectroscopic signature
that the same free-energy calculation retains the microscopic
quasiparticle rearrangement observed by STM.

\section{Calculation Results}

We first establish the method on the one defect for which the elementary pinning force has been measured directly, the Fe-site vacancy in FeSe, in which an Fe-site point defect was found by STM to pin a vortex through its core spectrum~\cite{Chen2024PRXPinning}.  For this configuration the calculated local force is \( f_p^{\rm loc}=1.18\times 10^{-4}\,\mathrm{N\,m^{-1}}\),
on the same scale as the microscopic STM estimate \(2.4\times10^{-4}\,\mathrm{N\,m^{-1}}\) and the \(0.8\times10^{-4}\,\mathrm{N\,m^{-1}}\) value inferred from transport~\cite{Chen2024PRXPinning}.

\begin{figure}[htbp]
\centering
\includegraphics[width=\columnwidth]{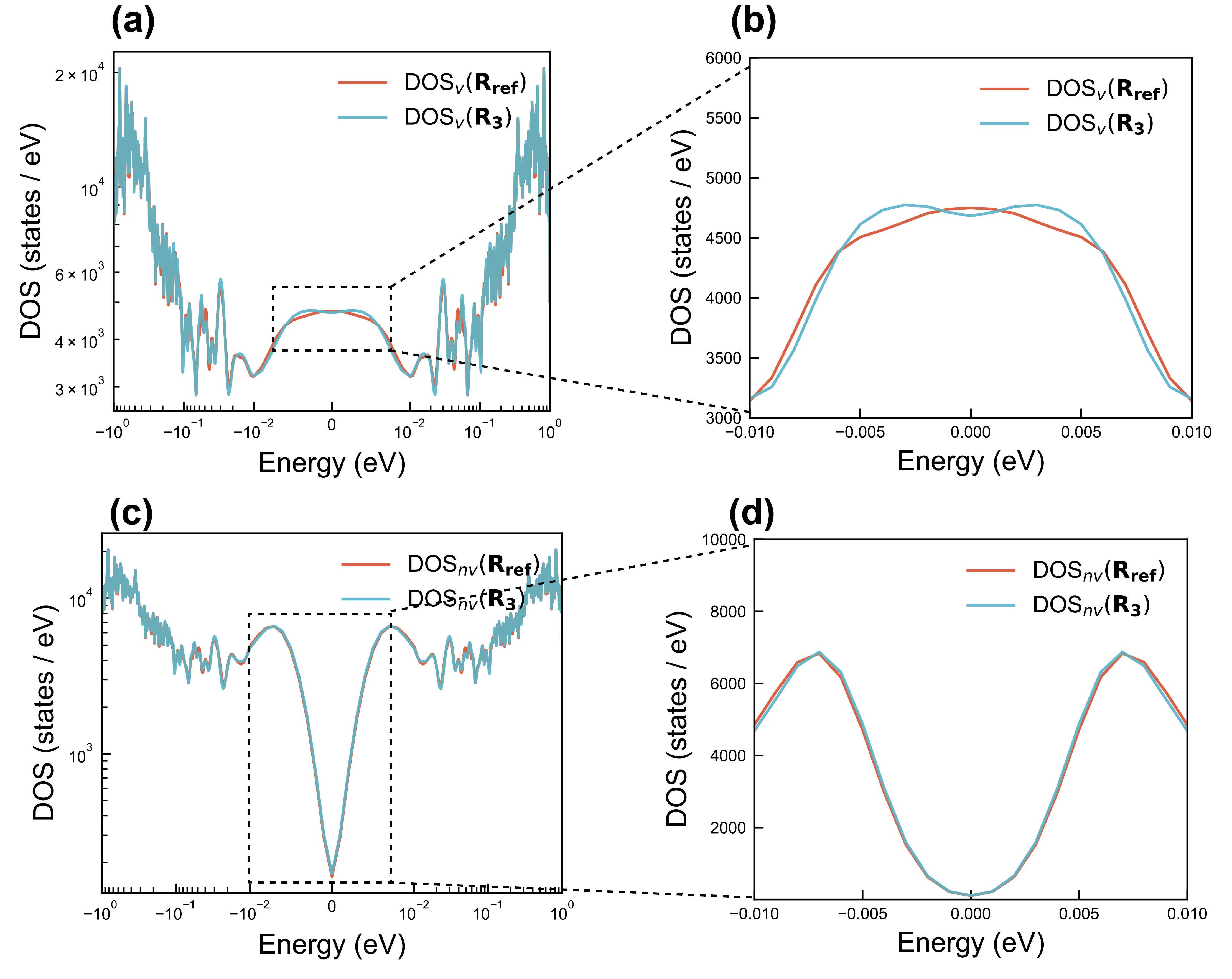}
\caption{Microscopic pinning mechanism driven by quasiparticle spectral reorganization. (a) Projected density of states (DOS) for the defect-on-core configuration $\mathbf R_3$ and the buffered reference $\mathbf R_{\rm ref}$, with the low-energy window enlarged in (b). When the defect is on the core, the zero-energy vortex-bound spectral weight is suppressed and redistributed to finite energies. For comparison, the corresponding no-vortex background states are shown in (c) and enlarged in (d), providing a stable baseline.}
\label{fig:mechanism_dos}
\end{figure}

The force scale alone is not the decisive test. Our central claim is that the defect pins a vortex by reorganizing the core quasiparticle spectrum, and this can be checked against the measured spectrum directly. Figure~\ref{fig:mechanism_dos}(a) shows the projected density of states (DOS) for the defect-on-core configuration $\mathbf R_3$ and the buffered reference $\mathbf R_{\rm ref}$, with the low-energy window enlarged in Fig.~\ref{fig:mechanism_dos}(b). When the defect is on the core, the zero-energy spectral weight is depleted and redistributed to finite energy, pushing the low-energy vortex-bound states off the Fermi level, exactly the elementary-pinning signature observed by STM~\cite{Chen2024PRXPinning}. For comparison, the corresponding no-vortex background states are shown in Figs.~\ref{fig:mechanism_dos}(c) and \ref{fig:mechanism_dos}(d), providing a stable baseline to verify this spectral shift. That this reorganization, and not the accompanying gap-amplitude suppression, controls the force follows from switching off the imposed defect-local suppression factor entirely: setting $s_a\equiv1$ changes the pinning energy from $-1.202$ to $-0.940$~meV and lowers the force from $1.183$ to $1.003\times10^{-4}\,\mathrm{N\,m^{-1}}$, so the pinning stays attractive and within the same order of magnitude, and the order-parameter suppression is not the controlling factor. The calculation thus reproduces the measured force scale and spectrum at once, confirming from first principles that the elementary pinning observed experimentally is a quasiparticle-spectral effect.

\begin{figure}[t]
\centering
\includegraphics[width=\columnwidth]{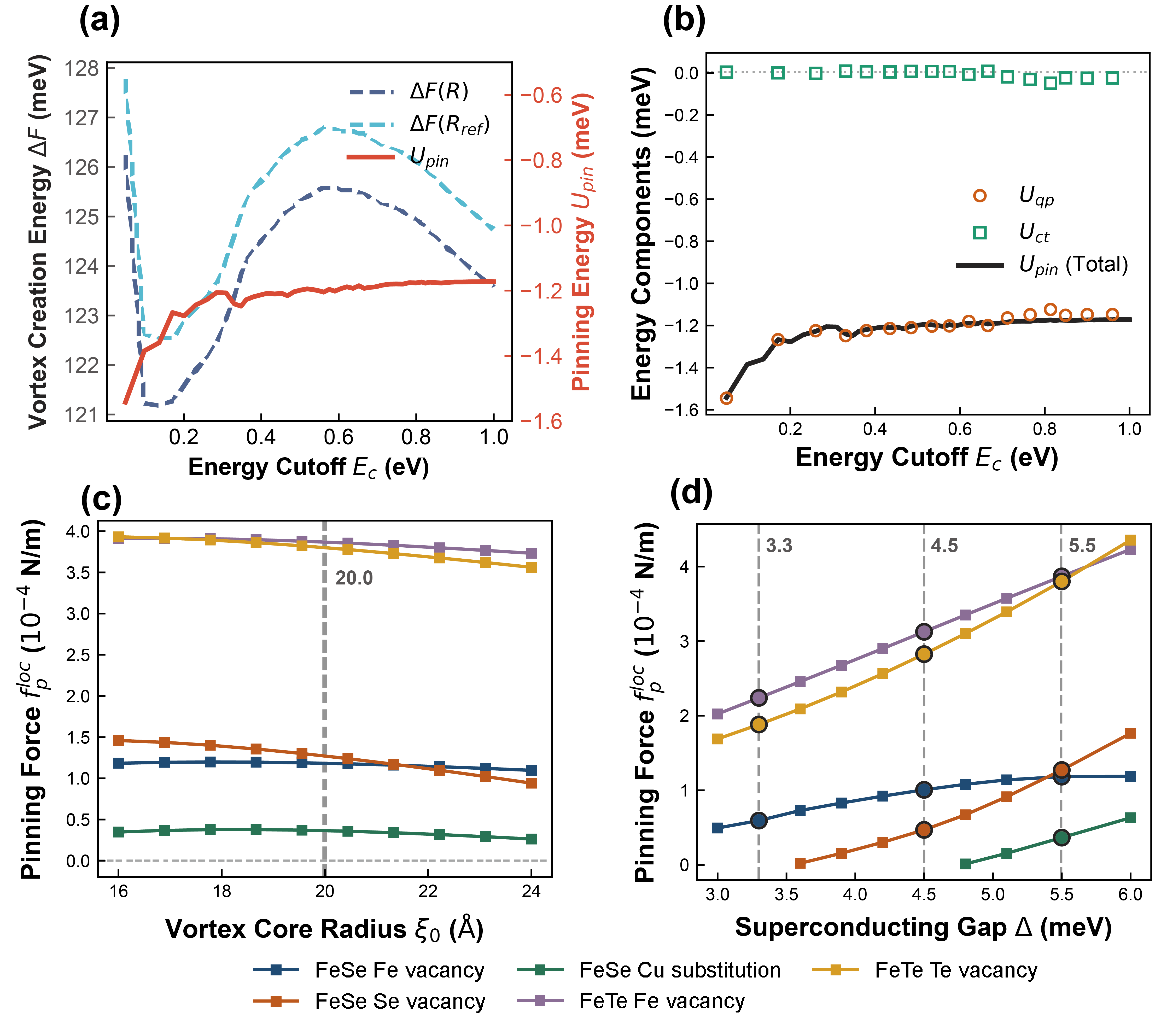}
\caption{Numerical robustness of the finite-box calculation and parametric dependence. (a) Vortex creation energies $\Delta F$ and pinning energy $U_{\rm pin}$ versus cutoff $E_c$. (b) Energy components $U_{\rm qp}$ and $U_{\rm ct}$ versus $E_c$, whose sum yields a converged total $U_{\rm pin}$. Extracted pinning force $f_p^{\rm loc}$ versus (c) vortex core radius $\xi_0$ and (d) superconducting gap $\Delta$. Vertical dashed lines mark the benchmark parameters $\xi_0=20$~\AA\ and $\Delta=5.5$~meV (the minimum of the anisotropic $\mathrm{(Li,Fe)OHFeSe}$ gap)~\cite{Chen2024PRXPinning} used in Table~\ref{tab:main_results}, alongside experimentally motivated FeTe gap scales $\Delta=3.3$ and $4.5$~meV~\cite{Qin2021FeTeBi2Te3,Ren2021OxygenFeTe,Yan2026StoichiometricFeTe}.}
\label{fig:cutoff_convergence}
\end{figure}

\begin{table*}[t]
\caption{
Finite-box pinning energies and local elementary forces for the five benchmark point defects.  The three energy columns use the buffered-reference difference \(U_{33-11}=\Delta F(\mathbf R_3)-\Delta F(\mathbf R_1)\), evaluated from Eq.~\eqref{eq:pinning_landscape}.  \(U_{\rm pin}^{\rm qp}\) and \(U_{\rm pin}^{\rm ct}\) are obtained by applying the same four-configuration subtraction to the quasiparticle trace and to the pairing counterterm in Eq.~\eqref{eq:bdg_free_energy}.  The force \(f_p^{\rm loc}\) is extracted separately from the nearest near-core difference \(U_{33-22}\) in Eq.~\eqref{eq:local_pinning_force}; it is not the maximum slope over all sampled segments.
}
\label{tab:main_results}
\begin{ruledtabular}
\begin{tabular}{lcccc}
Defect model &
\(U_{\rm pin}^{\rm qp}\) (meV) &
\(U_{\rm pin}^{\rm ct}\) (meV) &
\(U_{\rm pin}\) (meV) &
\(f_p^{\rm loc}\) \((10^{-4}\,\mathrm{N\,m^{-1}})\) \\
\hline
FeSe \(V_{\rm Fe}\)     & \(-1.2173\) & \(+0.0158\) & \(-1.2015\) & \(1.183\) \\
FeSe \(V_{\rm Se}\)     & \(-6.0167\) & \(+0.0360\) & \(-5.9807\) & \(1.271\) \\
FeTe \(V_{\rm Fe}\)     & \(-6.3235\) & \(-0.0252\) & \(-6.3487\) & \(3.868\) \\
FeTe \(V_{\rm Te}\)     & \(-8.6019\) & \(+0.0459\) & \(-8.5559\) & \(3.800\) \\
FeSe \(\mathrm{Cu}_{\rm Fe}\) & \(-1.3086\) & \(-0.0616\) & \(-1.3702\) & \(0.3650\) \\
\end{tabular}
\end{ruledtabular}
\end{table*}

This benchmark rests on a numerically stable free energy. The quasiparticle trace and the pairing counterterm each drift with the projected window $E_c$, yet their sum converges, with a residual high-window variation below $0.3$~meV [Figs.~\ref{fig:cutoff_convergence}(a) and \ref{fig:cutoff_convergence}(b)], the numerical counterpart of the cutoff cancellation established above and consistent with standard BCS renormalization~\cite{Bardeen1957BCS}. Perturbing the clean--defect boundary by the residual outer-shell Hamiltonian mismatch shifts $U_{\rm pin}$ only at the $10^{-2}$--$10^{-1}$~meV level, and reasonable variations of the vortex-core radius $\xi_0$ [Fig.~\ref{fig:cutoff_convergence}(c)] change the force slightly without altering its sign or scale. The FeSe Fe vacancy therefore fixes both the force and its microscopic origin, establishing the method.

\begin{figure}[htbp]
\centering
\includegraphics[width=\columnwidth]{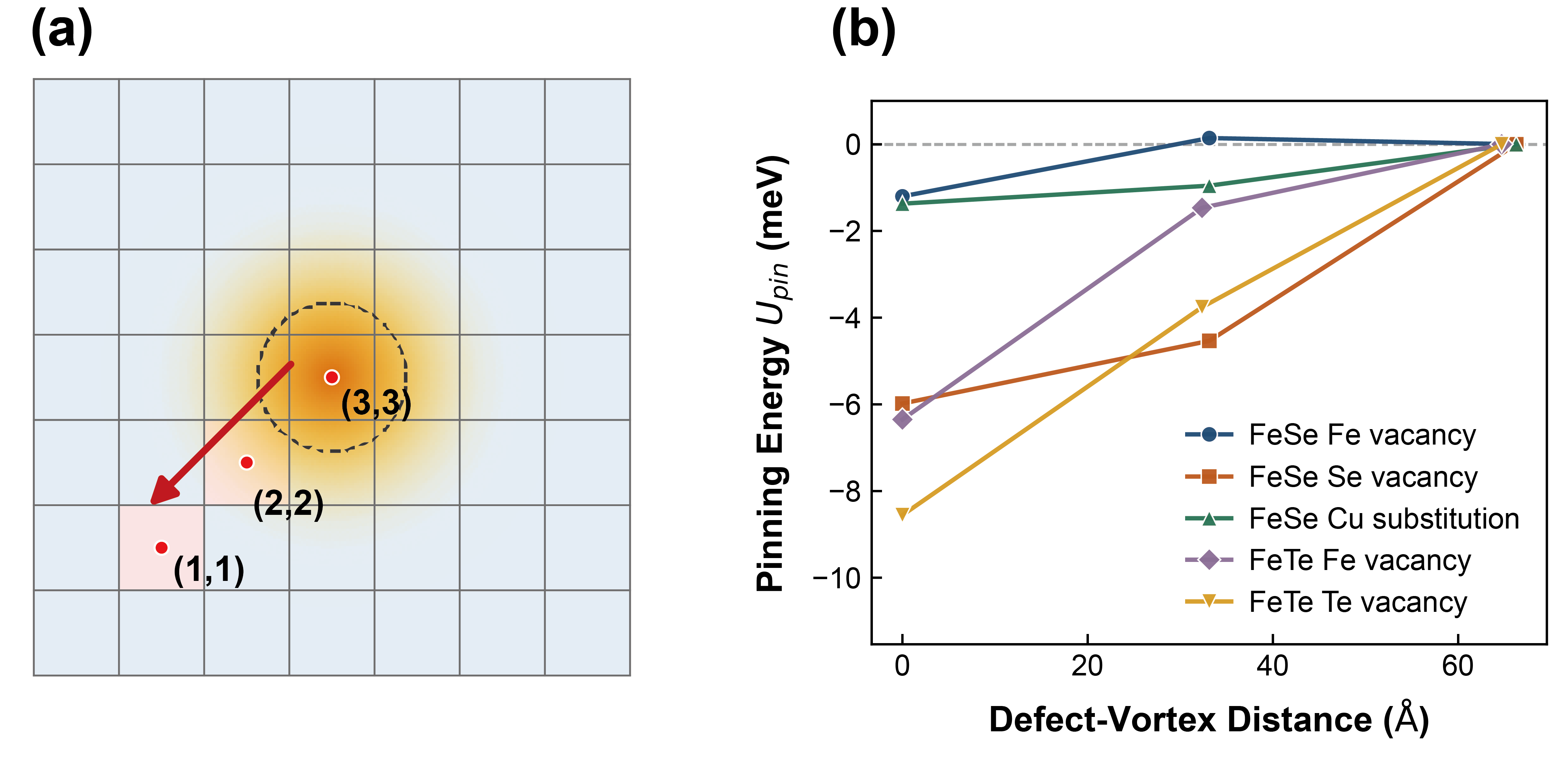}
\caption{(a) Schematic of the spatial scan used to extract the local pinning force. (b) Calculated pinning energy $U_{\rm pin}$ versus defect--vortex distance for benchmark point defects in FeSe and FeTe. Lines guide the eye through the four computed configurations. All defects have attractive center-to-reference binding; $f_p^{\rm loc}$ is the $\mathbf R_2$--$\mathbf R_3$ slope, while the steepest sampled segment for the FeSe Se vacancy lies between $\mathbf R_1$ and $\mathbf R_2$.}
\label{fig:pinning_landscape}
\end{figure}

We now apply the validated method to the full benchmark set, comprising Fe-site and chalcogen-site vacancies and Fe-site Cu substitution in FeSe and FeTe. As Fig.~\ref{fig:pinning_landscape} shows, every defect has a lower vortex-insertion energy at the defect-centered configuration \(\mathbf R_3\) than at the buffered reference \(\mathbf R_1\), and hence attractive binding in the convention of Eq.~\eqref{eq:pinning_landscape}. The landscape between the sampled positions need not, however, be monotonic: \(U_{33-11}\) measures the sampled well depth, whereas \(f_p^{\rm loc}\) measures only the near-center slope between \(\mathbf R_2\) and \(\mathbf R_3\). The energies and forces are collected in Table~\ref{tab:main_results}. The buffered-reference pinning energies span \(-1.20\) meV for the FeSe Fe vacancy to \(-8.56\) meV for the FeTe Te vacancy, the local forces are of order \(10^{-4}\,\mathrm{N\,m^{-1}}\), and FeTe vacancies give the largest local force at the common gap scale. In every case the quasiparticle trace is attractive, while the much smaller counterterm is a sign-dependent correction, so the net \(U_{\rm pin}\) remains attractive.

These results carry direct guidance for defect selection, as illustrated by the FeTe results. Superconductivity in FeTe has only recently been reported and remains under debate. It can be induced in thin films and interfaces such as one-unit-cell FeTe on $\mathrm{Bi_2Te_3}$ and oxygenated ultrathin FeTe on $\mathrm{SrTiO_3}$~\cite{Qin2021FeTeBi2Te3,Ren2021OxygenFeTe}, and stoichiometric FeTe freed of excess interstitial iron was very recently found to superconduct with a gap of about $4.5$~meV~\cite{Yan2026StoichiometricFeTe}. The STM/STS gap scales reported in these settings are of order a few meV. In the common benchmark calculation of Table~\ref{tab:main_results}, FeTe Fe and Te vacancies give the largest local forces, both about $3.8\times10^{-4}\,\mathrm{N\,m^{-1}}$; the Te vacancy gives the stronger buffered-reference pinning energy, whereas the Fe vacancy has the slightly larger near-core finite-difference force. While these benchmark values share the same $\Delta_0=5.5$~meV scale as the FeSe calculations, we also recomputed the FeTe vacancies using experimentally motivated FeTe gap scales~\cite{Qin2021FeTeBi2Te3,Ren2021OxygenFeTe}. As demonstrated in Fig.~\ref{fig:cutoff_convergence}(d), reducing the gap scale $\Delta_0$ lowers the absolute pinning energy and force, as expected, but both FeTe vacancies remain attractive and retain force scales of order $10^{-4}\,\mathrm{N\,m^{-1}}$. These results suggest that FeTe-based vacancy engineering is a promising target for experimental tests and demonstrate the potential of the present algorithm for high-throughput screening of stronger point pins.

The FeSe Se vacancy illustrates why the depth of the pinning well must be distinguished from the local slope of the landscape. Although the well is deep ($U_{33-11}=-5.98$~meV), most of the sampled energy decrease ($4.53$~meV, or $76\%$) occurs between $\mathbf R_1$ and $\mathbf R_2$, while the final approach from $\mathbf R_2$ to the vacancy-centered $\mathbf R_3$ lowers the energy by only $1.45$~meV. The near-core force $f_p^{\rm loc}$ is correspondingly modest, $1.27\times10^{-4}\,\mathrm{N\,m^{-1}}$, whereas the largest segment-averaged force in the scan, $3.99\times10^{-4}\,\mathrm{N\,m^{-1}}$, lies between $\mathbf R_1$ and $\mathbf R_2$. A site-resolved decomposition of the quasiparticle energy supports the same nonlocal picture: the attraction is not dominated by the shell nearest the vacancy but accumulates over the extended vortex-core halo. This nonlocality has a direct observable consequence. Under a finite Lorentz or intervortex driving force, the vortex settles where $\mathbf f_{\rm drive}+\mathbf f_p(\mathbf R)=0$; near depinning, this force-balanced position approaches the steepest part of the well rather than the vacancy site, and the present scan brackets that steepest segment between $\mathbf R_1$ and $\mathbf R_2$. A Se vacancy can therefore bind a vortex whose spectroscopic core center does not coincide with the vacancy, which may reconcile our result with the STM observation that a vortex core can overlap nearby Se vacancies without any detectable modification of the core states~\cite{Chen2024PRXPinning}. Since the four-point scan brackets only the interval of maximum slope, a finer two-dimensional map of the pinning landscape would be needed to locate the driven equilibrium position quantitatively.

\section{Conclusion}

We have given a first-principles, quantum-mechanical account of how a point
defect pins a vortex. Building the elementary force as a finite-box
vortex-insertion free energy, whose four-configuration subtraction isolates the
meV-scale interaction from much larger electronic-structure and finite-size
backgrounds, we show that the force is set by the defect-induced reorganization
of the vortex-core quasiparticle spectrum. This places elementary point-defect
pinning in the spectral content that survives beyond the Ginzburg--Landau
condensation-energy limit, to which the same free energy reduces only after its
quasiparticle spectrum is coarse-grained away.

Across five point defects in FeSe and FeTe the pinning is attractive; without
any fitting, the FeSe Fe vacancy matches the STM force scale and reproduces the
measured spectral reorganization, while FeTe vacancies pin most strongly.
Switching off the imposed gap suppression entirely lowers the force only within
the same order of magnitude, confirming the quasiparticle spectrum rather than
the order-parameter suppression as its controlling origin, while
internal-consistency and perturbation tests establish that the signal is
physical. The elementary force is not by itself the
sample-dependent \(J_c(B,T)\), but it is now a computable electronic-structure
quantity, turning the microscopic mechanism seen by STM into a predictive route
to the first-principles screening of defects for high-\(J_c\) superconductors.

\bmhead{Data availability}

No external datasets were used in this study. All experimental results presented in the manuscript can be fully generated, reproduced, and verified by running the source code archived on Zenodo at \url{https://doi.org/10.5281/zenodo.21790252}~\cite{ShiPinningCode}.

\bmhead{Code availability}

The source code implementing the finite-box projected Bogoliubov--de Gennes vortex-pinning framework used in this study is openly available. The exact version used to generate the results reported here is archived on Zenodo at \url{https://doi.org/10.5281/zenodo.21790252}~\cite{ShiPinningCode}, and the actively maintained version is hosted on GitHub at \url{https://github.com/de-Sitter/pinning_algorithm}. The code is distributed under an OSI-approved open-source license.

\begin{acknowledgments}
This work was supported by the National Natural Science Foundation of China (Grants No. 12188101, No. 12274081, No. 12225403, No. 92365302, No. 124B1003), National Key Research and Development Program of China (Grant No. 2024YFA1409800).
\end{acknowledgments}

\appendix

\section{First-principles implementation}
\label{app:implementation}

\begin{figure*}[t]
\centering
\includegraphics[width=\textwidth]{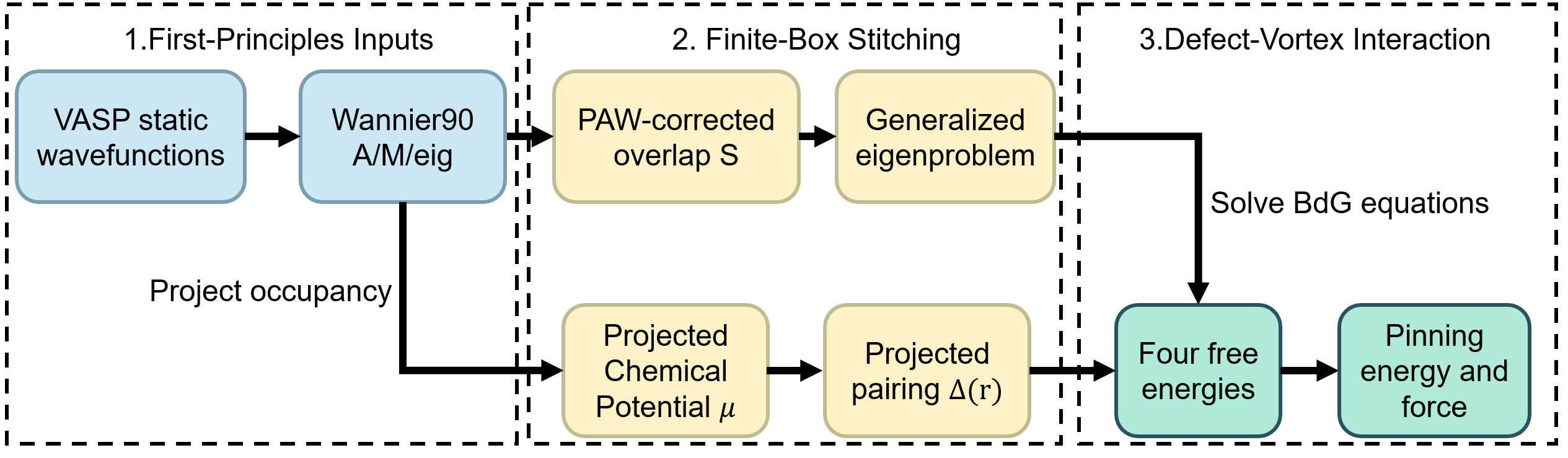}
\caption{Workflow of the first-principles elementary vortex-pinning calculation, organized in three stages. \textbf{(1) First-principles inputs:} static DFT wavefunctions of clean and defective supercells are wannierized into local Hamiltonians, and the DFT occupancies are projected to determine the finite-box electron count. \textbf{(2) Finite-box stitching:} embedding a defect patch in the clean background yields the PAW-corrected overlap matrix $S$ and the generalized eigenproblem [Eq.~\eqref{eq:generalized_eigenproblem}], together with the projected chemical potential $\mu$ and the projected pairing $\Delta(r)$ [Eq.~\eqref{eq:projected_pairing}]. \textbf{(3) Defect--vortex interaction:} solving the projected BdG equations [Eq.~\eqref{eq:projected_bdg}] gives the four free energies $F_{\rm v}$ and $F_{\rm nv}$ evaluated at $\mathbf R$ and $\mathbf R_{\rm ref}$ [Eq.~\eqref{eq:bdg_free_energy}], whose four-configuration combination yields the pinning energy $U_{\rm pin}$ [Eq.~\eqref{eq:pinning_landscape}] and the elementary force $f_p^{\rm loc}$ [Eq.~\eqref{eq:local_pinning_force}].}
\label{fig:algorithm_workflow}
\end{figure*}

This appendix details the first-principles construction of the normal-state and pairing ingredients that enter the projected BdG free energy of Eq.~\eqref{eq:bdg_free_energy}; Fig.~\ref{fig:algorithm_workflow} summarizes the pipeline. The normal-state input is generated from density functional theory~\cite{KohnSham1965DFT} with the PBE exchange-correlation functional~\cite{Perdew1996PBE}, as implemented in VASP~\cite{Kresse1996VASP}, for a clean supercell and for a supercell containing the point defect. For both we construct Wannier Hamiltonians for the same low-energy manifold of Fe-\(d\) and chalcogen-\(p\) states~\cite{MarzariVanderbilt1997Wannier,Mostofi2008Wannier90,Marzari2012WannierReview}.

Because the clean and defect Wannier functions come from independent calculations, their gauges are unrelated. We first establish the site and orbital correspondence by a Hungarian assignment~\cite{Kuhn1955Hungarian} followed by a unitary Procrustes alignment~\cite{Schonemann1966Procrustes}. After embedding the matched clean reference in the defect-patch indexing, we determine one dense unitary rotation \(Q\) for the entire defect basis by minimizing
\begin{equation}
\mathcal J(Q)=
\sum_{\mathbf T\in\mathcal R_H}
\left\|\left[Q^\dagger H_d(\mathbf T)Q-H_c(\mathbf T)\right]_{B,:}\right\|_F^2
,\qquad Q^\dagger Q=I.
\label{eq:global_gauge_objective}
\end{equation}
where \(B\) selects the outer-boundary Wannier rows and \(\mathcal R_H\) contains all 25 stored onsite and hopping blocks. The same fixed \(Q\) is used for every defect position in the normal-state construction, with
\begin{equation}
H_d'(\mathbf T)=Q^\dagger H_d(\mathbf T)Q,
\qquad
S_{dc}'=Q^\dagger S_{dc}.
\label{eq:global_basis_transform}
\end{equation}
Thus the rotation reduces the clean--defect boundary mismatch while preserving orthonormality and the isolated defect spectrum; it does not set the residual mismatch to zero.

A \emph{patch} is a \(6\times6\) block of the Wannier Hamiltonian; the finite box is a \(7\times7\) array of patches, one defect patch embedded in forty-eight clean patches. The defect patch is placed at lattice-registered origins; at the central position it is centered on the fixed vortex, the defect-centered configuration \(\mathbf R_3\), while the off-center positions supply the remaining separations, including the nearest one \(\mathbf R_2\). Inside the defect patch the onsite and hopping blocks are the globally rotated defect blocks, whereas the surrounding cells and the patch boundary take the clean blocks. The influence of the remaining boundary mismatch is checked a posteriori by the boundary-perturbation test.

Since the two Wannier bases are not mutually orthonormal, the stitched box carries a nontrivial overlap metric, and the clean--defect cross-overlap must be evaluated from the PAW all-electron inner product rather than assumed to vanish. From the PAW projector coefficients we form the raw-band overlap \(M_{dc}(\mathbf k)\) between clean and defect Bloch states and rotate it into the Wannier gauge,
\begin{equation}
S_{dc}(\mathbf k)=U_d^\dagger(\mathbf k)\,M_{dc}(\mathbf k)\,U_c(\mathbf k),
\label{eq:cross_overlap}
\end{equation}
with \(U_{c,d}\) the clean and defect gauge matrices. After the global rotation in Eq.~\eqref{eq:global_basis_transform}, the overlap matrix in the union basis of clean and defect orbitals is
\begin{equation}
S(\mathbf k)=
\begin{pmatrix}
I & S_{dc}'(\mathbf k)\\[2pt]
S_{dc}^{\prime\dagger}(\mathbf k) & I
\end{pmatrix},
\label{eq:union_overlap}
\end{equation}
its diagonal blocks the identity because each Wannier set is internally orthonormal; Fourier transforming gives the real-space metric \(S(\mathbf R)\). For each defect position \(\mathbf R\) we then assemble the normal-state Hamiltonian \(H_0(\mathbf R)\), including this PAW-corrected clean--defect cross overlap~\cite{Blochl1994PAW,KresseJoubert1999PAW}, and obtain the normal-state eigenstates from the generalized eigenproblem
\begin{equation}
\begin{aligned}
    &H_0(\mathbf R) C(\mathbf R)
=
S(\mathbf R) C(\mathbf R)\varepsilon(\mathbf R),\\
&\qquad
C^\dagger(\mathbf R)S(\mathbf R)C(\mathbf R)=1 .
\end{aligned}
\label{eq:generalized_eigenproblem}
\end{equation}
Treating the stitched basis as orthonormal (\(S=I\)) would misrepresent the low-energy spectrum on which the BdG projection rests.

The chemical potential is fixed inside the Wannier subspace rather than taken from the raw DFT Fermi level. We project the DFT occupations onto the clean and defect subspaces to obtain the per-patch electron counts \(N_W^{\rm clean}\) and \(N_W^{\rm defect}\), and hold the total box filling at \(N_{\rm target}=48\,N_W^{\rm clean}+N_W^{\rm defect}\); for each origin \(\mu(\mathbf R)\) is adjusted to reproduce this projected occupancy.

The superconducting state is introduced as a calibrated BdG background. For an orbital \(a\) centered at \(\mathbf r_a\), with vortex center \(\mathbf r_v\), we use
\begin{equation}
\begin{aligned}
    &\Delta_a^{(\rm v)}(\mathbf R)
=
\Delta_0\,s_a(\mathbf R)
\tanh\!\left(
\frac{|\mathbf r_a-\mathbf r_v|}{\xi_0}
\right)
e^{i\theta_a},\\
&\qquad
\Delta_a^{(\rm nv)}(\mathbf R)
=
\Delta_0\,s_a(\mathbf R),
\end{aligned}
\label{eq:pairing_texture}
\end{equation}
where \(\theta_a\) is the polar angle of \(\mathbf r_a-\mathbf r_v\), so the vortex texture carries one phase winding and a core suppression on the scale \(\xi_0\)~\cite{Clem1975VortexCore}, while the no-vortex reference keeps the same gap amplitude and the same defect-local background. The factor \(s_a(\mathbf R)\) is a smooth, non-negative defect-local amplitude factor applied identically to the vortex and no-vortex references,
\begin{equation}
s_a(\mathbf R)=\max\!\left[\,0,\ 1-\eta\,e^{-|\mathbf r_a-\mathbf r_d|^2/\ell_{\rm imp}^2}\,\right],
\label{eq:suppression}
\end{equation}
with \(\mathbf r_d\) the defect position, \(\eta\) the suppression strength, and \(\ell_{\rm imp}\) its range. The gap scale \(\Delta_0\) is an experimentally calibrated material input~\cite{Chen2024PRXPinning}; a self-consistent determination, for example within superconducting density-functional theory~\cite{OliveiraGrossKohn1988SCDFT}, could replace it, but predicting the pairing scale is not the aim here.

For the dense defect-patch pairing matrix, the profiles in Eq.~\eqref{eq:pairing_texture} are used as continuous fields by replacing \(\mathbf r_a\) with \(\mathbf r\). We first integrate them in the localized Wannier basis and then add the PAW one-centre correction~\cite{Blochl1994PAW,KresseJoubert1999PAW}:
\begin{equation}
\begin{aligned}
\Delta_{ij}^{\rm loc,(x)}
&=\int_{\rm BvK}d^3r\,
w_i^*(\mathbf r)\Delta^{(x)}(\mathbf r)w_j^*(\mathbf r),\\
\Delta_{\rm AE}^{(x)}
&=\Delta_{\rm smooth,BvK}^{(x)}
+\sum_{\mathbf T,a}P_{a\mathbf T}^\dagger
\left(D_{\rm AE}^{a,(x)}-D_{\rm PS}^{a,(x)}\right)
P_{a\mathbf T}^*,\\
\Delta_d^{\prime(x)}
&=Q^\dagger\Delta_{\rm AE}^{(x)}Q^*,
\qquad x={\rm v},{\rm nv}.
\end{aligned}
\label{eq:paw_pairing_matrix}
\end{equation}
The first line defines the matrix in the localized Wannier basis. In the PAW reconstruction on the second line, \(P_{a\mathbf T}\) contains the projector coefficients, \(D_{\rm AE/PS}^{a,(x)}\) are the all-electron/pseudo one-centre matrix elements of the same pairing field, and the sum covers all atoms and the complete \(3\times3\times1\) Born--von Karman translation domain. The final line applies the same fixed global rotation to both vortex sectors after the pairing matrix has been constructed. The transformed dense matrix \(\Delta_d^{\prime(x)}\) is used inside the defect patch, the clean exterior retains the diagonal local-field approximation, and the defect--clean cross-boundary pairing blocks are zero.

The BdG subspace is then selected by a common physical energy window,
\begin{equation}
\mathcal C(\mathbf R;E_c)
=
\left\{
i:\,
|\varepsilon_i(\mathbf R)-\mu(\mathbf R)|\le E_c
\right\},
\label{eq:common_cutoff}
\end{equation}
which compares the same physical energy range at every defect position rather than a fixed number of retained states.

The pairing matrix is projected with the same generalized normal-state eigenvectors,
\begin{equation}
\Delta_{\mathcal C}^{(x)}(\mathbf R)
=
C_{\mathcal C}^{\dagger}(\mathbf R)
\Delta^{(x)}(\mathbf R)
C_{\mathcal C}^{*}(\mathbf R),
\qquad
x={\rm v},{\rm nv},
\label{eq:projected_pairing}
\end{equation}
where \(C_{\mathcal C}\) collects the columns selected by Eq.~\eqref{eq:common_cutoff}. The finite-box \(\Delta^{(x)}(\mathbf R)\) is assembled from the transformed dense defect-patch block \(\Delta_d^{\prime(x)}\), the diagonal local-field matrix in the clean exterior, and zero defect--clean cross-boundary blocks. Within this subspace the one-block spin-singlet BdG Hamiltonian~\cite{deGennes1966Superconductivity} is
\begin{equation}
\begin{aligned}
    &\mathcal H_{\rm BdG}^{(x)}(\mathbf R)
=
\begin{pmatrix}
\xi_{\mathcal C}(\mathbf R) & \Delta_{\mathcal C}^{(x)}(\mathbf R)\\
\Delta_{\mathcal C}^{(x)\dagger}(\mathbf R) & -\xi_{\mathcal C}(\mathbf R)
\end{pmatrix},\\
&\qquad
\xi_i(\mathbf R)=\varepsilon_i(\mathbf R)-\mu(\mathbf R).
\end{aligned}
\label{eq:projected_bdg}
\end{equation}
The pairing coupling \(g(E_c)\) is calibrated on the clean system through the gap equation at the chosen \(\Delta_0\) and cutoff, so that the quasiparticle trace and the counterterm in Eq.~\eqref{eq:bdg_free_energy} are evaluated consistently in the same window. Collecting these ingredients, the projected BdG free energy is
\begin{equation}
F_x(\mathbf R;E_c)
=
-\tfrac12\,{\rm Tr}\left|\mathcal H_{\rm BdG}^{(x)}(\mathbf R;E_c)\right|
+
\frac{1}{g(E_c)}\left\|\Delta_{\mathcal C}^{(x)}(\mathbf R)\right\|_F^2 ,
\tag{\ref{eq:bdg_free_energy}}
\end{equation}
the quasiparticle spectral trace plus the cutoff-consistent counterterm of Eq.~\eqref{eq:bdg_free_energy}. Evaluating it for the four configurations \(x={\rm v},{\rm nv}\) at \(\mathbf R\) and \(\mathbf R_{\rm ref}\), and substituting \(F_{\rm v}\) and \(F_{\rm nv}\) into the pinning-energy definition Eq.~\eqref{eq:pinning_landscape}, yields the finite-box pinning landscape.

\bibliography{ref}

\begin{thebibliography}{37}%
\makeatletter
\providecommand \@ifxundefined [1]{%
 \@ifx{#1\undefined}
}%
\providecommand \@ifnum [1]{%
 \ifnum #1\expandafter \@firstoftwo
 \else \expandafter \@secondoftwo
 \fi
}%
\providecommand \@ifx [1]{%
 \ifx #1\expandafter \@firstoftwo
 \else \expandafter \@secondoftwo
 \fi
}%
\providecommand \natexlab [1]{#1}%
\providecommand \enquote  [1]{``#1''}%
\providecommand \bibnamefont  [1]{#1}%
\providecommand \bibfnamefont [1]{#1}%
\providecommand \citenamefont [1]{#1}%
\providecommand \href@noop [0]{\@secondoftwo}%
\providecommand \href [0]{\begingroup \@sanitize@url \@href}%
\providecommand \@href[1]{\@@startlink{#1}\@@href}%
\providecommand \@@href[1]{\endgroup#1\@@endlink}%
\providecommand \@sanitize@url [0]{\catcode `\\12\catcode `\$12\catcode
  `\&12\catcode `\#12\catcode `\^12\catcode `\_12\catcode `\%12\relax}%
\providecommand \@@startlink[1]{}%
\providecommand \@@endlink[0]{}%
\providecommand \url  [0]{\begingroup\@sanitize@url \@url }%
\providecommand \@url [1]{\endgroup\@href {#1}{\urlprefix }}%
\providecommand \urlprefix  [0]{URL }%
\providecommand \Eprint [0]{\href }%
\providecommand \doibase [0]{https://doi.org/}%
\providecommand \selectlanguage [0]{\@gobble}%
\providecommand \bibinfo  [0]{\@secondoftwo}%
\providecommand \bibfield  [0]{\@secondoftwo}%
\providecommand \translation [1]{[#1]}%
\providecommand \BibitemOpen [0]{}%
\providecommand \bibitemStop [0]{}%
\providecommand \bibitemNoStop [0]{.\EOS\space}%
\providecommand \EOS [0]{\spacefactor3000\relax}%
\providecommand \BibitemShut  [1]{\csname bibitem#1\endcsname}%
\let\auto@bib@innerbib\@empty
\bibitem [{\citenamefont {Larbalestier}\ \emph {et~al.}(2001)\citenamefont
  {Larbalestier}, \citenamefont {Gurevich}, \citenamefont {Feldmann},\ and\
  \citenamefont {Polyanskii}}]{Larbalestier2001HighTc}%
  \BibitemOpen
  \bibfield  {author} {\bibinfo {author} {\bibfnamefont {D.}~\bibnamefont
  {Larbalestier}}, \bibinfo {author} {\bibfnamefont {A.}~\bibnamefont
  {Gurevich}}, \bibinfo {author} {\bibfnamefont {D.~M.}\ \bibnamefont
  {Feldmann}},\ and\ \bibinfo {author} {\bibfnamefont {A.}~\bibnamefont
  {Polyanskii}},\ }\bibfield  {title} {\bibinfo {title} {High-{$T_c$}
  superconducting materials for electric power applications},\ }\href
  {https://doi.org/10.1038/35104654} {\bibfield  {journal} {\bibinfo  {journal}
  {Nature}\ }\textbf {\bibinfo {volume} {414}},\ \bibinfo {pages} {368}
  (\bibinfo {year} {2001})}\BibitemShut {NoStop}%
\bibitem [{\citenamefont {Abrikosov}(1957)}]{Abrikosov1957TypeII}%
  \BibitemOpen
  \bibfield  {author} {\bibinfo {author} {\bibfnamefont {A.~A.}\ \bibnamefont
  {Abrikosov}},\ }\bibfield  {title} {\bibinfo {title} {On the magnetic
  properties of superconductors of the second group},\ }\href
  {https://jetp.ras.ru/cgi-bin/e/index/e/5/6/p1174?a=list} {\bibfield
  {journal} {\bibinfo  {journal} {Sov. Phys. JETP}\ }\textbf {\bibinfo {volume}
  {5}},\ \bibinfo {pages} {1174} (\bibinfo {year} {1957})},\ \bibinfo {note}
  {russian original: Zh. Eksp. Teor. Fiz. 32, 1442 (1957)}\BibitemShut
  {NoStop}%
\bibitem [{\citenamefont {Bean}(1964)}]{Bean1964CriticalState}%
  \BibitemOpen
  \bibfield  {author} {\bibinfo {author} {\bibfnamefont {C.~P.}\ \bibnamefont
  {Bean}},\ }\bibfield  {title} {\bibinfo {title} {Magnetization of high-field
  superconductors},\ }\href {https://doi.org/10.1103/RevModPhys.36.31}
  {\bibfield  {journal} {\bibinfo  {journal} {Rev. Mod. Phys.}\ }\textbf
  {\bibinfo {volume} {36}},\ \bibinfo {pages} {31} (\bibinfo {year}
  {1964})}\BibitemShut {NoStop}%
\bibitem [{\citenamefont {Anderson}\ and\ \citenamefont
  {Kim}(1964)}]{AndersonKim1964FluxCreep}%
  \BibitemOpen
  \bibfield  {author} {\bibinfo {author} {\bibfnamefont {P.~W.}\ \bibnamefont
  {Anderson}}\ and\ \bibinfo {author} {\bibfnamefont {Y.~B.}\ \bibnamefont
  {Kim}},\ }\bibfield  {title} {\bibinfo {title} {Hard superconductivity:
  Theory of the motion of abrikosov flux lines},\ }\href
  {https://doi.org/10.1103/RevModPhys.36.39} {\bibfield  {journal} {\bibinfo
  {journal} {Rev. Mod. Phys.}\ }\textbf {\bibinfo {volume} {36}},\ \bibinfo
  {pages} {39} (\bibinfo {year} {1964})}\BibitemShut {NoStop}%
\bibitem [{\citenamefont {Dew-Hughes}(1974)}]{DewHughes1974FluxPinning}%
  \BibitemOpen
  \bibfield  {author} {\bibinfo {author} {\bibfnamefont {D.}~\bibnamefont
  {Dew-Hughes}},\ }\bibfield  {title} {\bibinfo {title} {Flux pinning
  mechanisms in type ii superconductors},\ }\href
  {https://doi.org/10.1080/14786439808206556} {\bibfield  {journal} {\bibinfo
  {journal} {Philos. Mag.}\ }\textbf {\bibinfo {volume} {30}},\ \bibinfo
  {pages} {293} (\bibinfo {year} {1974})}\BibitemShut {NoStop}%
\bibitem [{\citenamefont {Larkin}\ and\ \citenamefont
  {Ovchinnikov}(1979)}]{LarkinOvchinnikov1979CollectivePinning}%
  \BibitemOpen
  \bibfield  {author} {\bibinfo {author} {\bibfnamefont {A.~I.}\ \bibnamefont
  {Larkin}}\ and\ \bibinfo {author} {\bibfnamefont {Y.~N.}\ \bibnamefont
  {Ovchinnikov}},\ }\bibfield  {title} {\bibinfo {title} {Pinning in type ii
  superconductors},\ }\href {https://doi.org/10.1007/BF00117160} {\bibfield
  {journal} {\bibinfo  {journal} {J. Low Temp. Phys.}\ }\textbf {\bibinfo
  {volume} {34}},\ \bibinfo {pages} {409} (\bibinfo {year} {1979})}\BibitemShut
  {NoStop}%
\bibitem [{\citenamefont {Blatter}\ \emph {et~al.}(1994)\citenamefont
  {Blatter}, \citenamefont {Feigel'man}, \citenamefont {Geshkenbein},
  \citenamefont {Larkin},\ and\ \citenamefont {Vinokur}}]{Blatter1994Vortices}%
  \BibitemOpen
  \bibfield  {author} {\bibinfo {author} {\bibfnamefont {G.}~\bibnamefont
  {Blatter}}, \bibinfo {author} {\bibfnamefont {M.~V.}\ \bibnamefont
  {Feigel'man}}, \bibinfo {author} {\bibfnamefont {V.~B.}\ \bibnamefont
  {Geshkenbein}}, \bibinfo {author} {\bibfnamefont {A.~I.}\ \bibnamefont
  {Larkin}},\ and\ \bibinfo {author} {\bibfnamefont {V.~M.}\ \bibnamefont
  {Vinokur}},\ }\bibfield  {title} {\bibinfo {title} {Vortices in
  high-temperature superconductors},\ }\href
  {https://doi.org/10.1103/RevModPhys.66.1125} {\bibfield  {journal} {\bibinfo
  {journal} {Rev. Mod. Phys.}\ }\textbf {\bibinfo {volume} {66}},\ \bibinfo
  {pages} {1125} (\bibinfo {year} {1994})}\BibitemShut {NoStop}%
\bibitem [{\citenamefont {Thuneberg}\ \emph {et~al.}(1984)\citenamefont
  {Thuneberg}, \citenamefont {Kurkij{\"a}rvi},\ and\ \citenamefont
  {Rainer}}]{Thuneberg1984ElementaryPinning}%
  \BibitemOpen
  \bibfield  {author} {\bibinfo {author} {\bibfnamefont {E.~V.}\ \bibnamefont
  {Thuneberg}}, \bibinfo {author} {\bibfnamefont {J.}~\bibnamefont
  {Kurkij{\"a}rvi}},\ and\ \bibinfo {author} {\bibfnamefont {D.}~\bibnamefont
  {Rainer}},\ }\bibfield  {title} {\bibinfo {title} {Elementary-flux-pinning
  potential in type-ii superconductors},\ }\href
  {https://doi.org/10.1103/PhysRevB.29.3913} {\bibfield  {journal} {\bibinfo
  {journal} {Phys. Rev. B}\ }\textbf {\bibinfo {volume} {29}},\ \bibinfo
  {pages} {3913} (\bibinfo {year} {1984})}\BibitemShut {NoStop}%
\bibitem [{\citenamefont {Hyun}\ \emph {et~al.}(1987)\citenamefont {Hyun},
  \citenamefont {Finnemore}, \citenamefont {Schwartzkopf},\ and\ \citenamefont
  {Clem}}]{Hyun1987ElementaryForce}%
  \BibitemOpen
  \bibfield  {author} {\bibinfo {author} {\bibfnamefont {O.~B.}\ \bibnamefont
  {Hyun}}, \bibinfo {author} {\bibfnamefont {D.~K.}\ \bibnamefont {Finnemore}},
  \bibinfo {author} {\bibfnamefont {L.}~\bibnamefont {Schwartzkopf}},\ and\
  \bibinfo {author} {\bibfnamefont {J.~R.}\ \bibnamefont {Clem}},\ }\bibfield
  {title} {\bibinfo {title} {Elementary pinning force for a superconducting
  vortex},\ }\href {https://doi.org/10.1103/PhysRevLett.58.599} {\bibfield
  {journal} {\bibinfo  {journal} {Phys. Rev. Lett.}\ }\textbf {\bibinfo
  {volume} {58}},\ \bibinfo {pages} {599} (\bibinfo {year} {1987})}\BibitemShut
  {NoStop}%
\bibitem [{\citenamefont {Caroli}\ \emph {et~al.}(1964)\citenamefont {Caroli},
  \citenamefont {de~Gennes},\ and\ \citenamefont {Matricon}}]{Caroli1964CdGM}%
  \BibitemOpen
  \bibfield  {author} {\bibinfo {author} {\bibfnamefont {C.}~\bibnamefont
  {Caroli}}, \bibinfo {author} {\bibfnamefont {P.~G.}\ \bibnamefont
  {de~Gennes}},\ and\ \bibinfo {author} {\bibfnamefont {J.}~\bibnamefont
  {Matricon}},\ }\bibfield  {title} {\bibinfo {title} {Bound fermion states on
  a vortex line in a type ii superconductor},\ }\href
  {https://doi.org/10.1016/0031-9163(64)90375-0} {\bibfield  {journal}
  {\bibinfo  {journal} {Phys. Lett.}\ }\textbf {\bibinfo {volume} {9}},\
  \bibinfo {pages} {307} (\bibinfo {year} {1964})}\BibitemShut {NoStop}%
\bibitem [{\citenamefont {Hess}\ \emph {et~al.}(1989)\citenamefont {Hess},
  \citenamefont {Robinson}, \citenamefont {Dynes}, \citenamefont {Valles},\
  and\ \citenamefont {Waszczak}}]{Hess1989STMFluxLattice}%
  \BibitemOpen
  \bibfield  {author} {\bibinfo {author} {\bibfnamefont {H.~F.}\ \bibnamefont
  {Hess}}, \bibinfo {author} {\bibfnamefont {R.~B.}\ \bibnamefont {Robinson}},
  \bibinfo {author} {\bibfnamefont {R.~C.}\ \bibnamefont {Dynes}}, \bibinfo
  {author} {\bibfnamefont {J.~M.}\ \bibnamefont {Valles}, \bibfnamefont
  {Jr.}},\ and\ \bibinfo {author} {\bibfnamefont {J.~V.}\ \bibnamefont
  {Waszczak}},\ }\bibfield  {title} {\bibinfo {title}
  {Scanning-tunneling-microscope observation of the abrikosov flux lattice and
  the density of states near and inside a fluxoid},\ }\href
  {https://doi.org/10.1103/PhysRevLett.62.214} {\bibfield  {journal} {\bibinfo
  {journal} {Phys. Rev. Lett.}\ }\textbf {\bibinfo {volume} {62}},\ \bibinfo
  {pages} {214} (\bibinfo {year} {1989})}\BibitemShut {NoStop}%
\bibitem [{\citenamefont {Chen}\ \emph {et~al.}(2020)\citenamefont {Chen},
  \citenamefont {Liu}, \citenamefont {Bao}, \citenamefont {Yan}, \citenamefont
  {Wang}, \citenamefont {Zhang},\ and\ \citenamefont
  {Feng}}]{Chen2020FeSeCdGM}%
  \BibitemOpen
  \bibfield  {author} {\bibinfo {author} {\bibfnamefont {C.}~\bibnamefont
  {Chen}}, \bibinfo {author} {\bibfnamefont {Q.}~\bibnamefont {Liu}}, \bibinfo
  {author} {\bibfnamefont {W.-C.}\ \bibnamefont {Bao}}, \bibinfo {author}
  {\bibfnamefont {Y.}~\bibnamefont {Yan}}, \bibinfo {author} {\bibfnamefont
  {Q.-H.}\ \bibnamefont {Wang}}, \bibinfo {author} {\bibfnamefont
  {T.}~\bibnamefont {Zhang}},\ and\ \bibinfo {author} {\bibfnamefont
  {D.}~\bibnamefont {Feng}},\ }\bibfield  {title} {\bibinfo {title}
  {Observation of discrete conventional caroli--de gennes--matricon states in
  the vortex core of single-layer {FeSe}/{SrTiO3}},\ }\href
  {https://doi.org/10.1103/PhysRevLett.124.097001} {\bibfield  {journal}
  {\bibinfo  {journal} {Phys. Rev. Lett.}\ }\textbf {\bibinfo {volume} {124}},\
  \bibinfo {pages} {097001} (\bibinfo {year} {2020})}\BibitemShut {NoStop}%
\bibitem [{\citenamefont {Zhang}\ \emph {et~al.}(2021)\citenamefont {Zhang},
  \citenamefont {Bao}, \citenamefont {Chen}, \citenamefont {Li}, \citenamefont
  {Lu}, \citenamefont {Hu}, \citenamefont {Yang}, \citenamefont {Zhao},
  \citenamefont {Yan}, \citenamefont {Dong}, \citenamefont {Wang},
  \citenamefont {Zhang},\ and\ \citenamefont
  {Feng}}]{Zhang2021LiFeOHFeSeVortexModes}%
  \BibitemOpen
  \bibfield  {author} {\bibinfo {author} {\bibfnamefont {T.~Z.}\ \bibnamefont
  {Zhang}}, \bibinfo {author} {\bibfnamefont {W.~C.}\ \bibnamefont {Bao}},
  \bibinfo {author} {\bibfnamefont {C.}~\bibnamefont {Chen}}, \bibinfo {author}
  {\bibfnamefont {D.}~\bibnamefont {Li}}, \bibinfo {author} {\bibfnamefont
  {Z.~Y.~W.}\ \bibnamefont {Lu}}, \bibinfo {author} {\bibfnamefont {Y.~N.}\
  \bibnamefont {Hu}}, \bibinfo {author} {\bibfnamefont {W.~T.}\ \bibnamefont
  {Yang}}, \bibinfo {author} {\bibfnamefont {D.~M.}\ \bibnamefont {Zhao}},
  \bibinfo {author} {\bibfnamefont {Y.~J.}\ \bibnamefont {Yan}}, \bibinfo
  {author} {\bibfnamefont {X.~L.}\ \bibnamefont {Dong}}, \bibinfo {author}
  {\bibfnamefont {Q.-H.}\ \bibnamefont {Wang}}, \bibinfo {author}
  {\bibfnamefont {T.}~\bibnamefont {Zhang}},\ and\ \bibinfo {author}
  {\bibfnamefont {D.~L.}\ \bibnamefont {Feng}},\ }\bibfield  {title} {\bibinfo
  {title} {Observation of distinct spatial distributions of the zero and
  nonzero energy vortex modes in
  {$(\mathrm{Li}_{0.84}\mathrm{Fe}_{0.16})\mathrm{OHFeSe}$}},\ }\href
  {https://doi.org/10.1103/PhysRevLett.126.127001} {\bibfield  {journal}
  {\bibinfo  {journal} {Phys. Rev. Lett.}\ }\textbf {\bibinfo {volume} {126}},\
  \bibinfo {pages} {127001} (\bibinfo {year} {2021})}\BibitemShut {NoStop}%
\bibitem [{\citenamefont {Zhang}\ \emph {et~al.}(2023)\citenamefont {Zhang},
  \citenamefont {Hu}, \citenamefont {Su}, \citenamefont {Chen}, \citenamefont
  {Wang}, \citenamefont {Li}, \citenamefont {Lu}, \citenamefont {Yang},
  \citenamefont {Zhang}, \citenamefont {Dong}, \citenamefont {Wang},
  \citenamefont {Wang}, \citenamefont {Feng},\ and\ \citenamefont
  {Zhang}}]{Zhang2023LiFeOHFeSeImpurity}%
  \BibitemOpen
  \bibfield  {author} {\bibinfo {author} {\bibfnamefont {T.~Z.}\ \bibnamefont
  {Zhang}}, \bibinfo {author} {\bibfnamefont {Y.~N.}\ \bibnamefont {Hu}},
  \bibinfo {author} {\bibfnamefont {W.}~\bibnamefont {Su}}, \bibinfo {author}
  {\bibfnamefont {C.}~\bibnamefont {Chen}}, \bibinfo {author} {\bibfnamefont
  {X.}~\bibnamefont {Wang}}, \bibinfo {author} {\bibfnamefont {D.}~\bibnamefont
  {Li}}, \bibinfo {author} {\bibfnamefont {Z.~Y.~W.}\ \bibnamefont {Lu}},
  \bibinfo {author} {\bibfnamefont {W.~T.}\ \bibnamefont {Yang}}, \bibinfo
  {author} {\bibfnamefont {Q.~L.}\ \bibnamefont {Zhang}}, \bibinfo {author}
  {\bibfnamefont {X.~L.}\ \bibnamefont {Dong}}, \bibinfo {author}
  {\bibfnamefont {R.}~\bibnamefont {Wang}}, \bibinfo {author} {\bibfnamefont
  {X.~Q.}\ \bibnamefont {Wang}}, \bibinfo {author} {\bibfnamefont {D.~L.}\
  \bibnamefont {Feng}},\ and\ \bibinfo {author} {\bibfnamefont
  {T.}~\bibnamefont {Zhang}},\ }\bibfield  {title} {\bibinfo {title} {Phase
  shift and magnetic anisotropy induced field splitting of impurity states in
  {$(\mathrm{Li}_{1-x}\mathrm{Fe}_x)\mathrm{OHFeSe}$} superconductor},\ }\href
  {https://doi.org/10.1103/PhysRevLett.130.206001} {\bibfield  {journal}
  {\bibinfo  {journal} {Phys. Rev. Lett.}\ }\textbf {\bibinfo {volume} {130}},\
  \bibinfo {pages} {206001} (\bibinfo {year} {2023})}\BibitemShut {NoStop}%
\bibitem [{\citenamefont {Chen}\ \emph {et~al.}(2024)\citenamefont {Chen},
  \citenamefont {Liu}, \citenamefont {Chen}, \citenamefont {Hu}, \citenamefont
  {Zhang}, \citenamefont {Li}, \citenamefont {Wang}, \citenamefont {Wang},
  \citenamefont {Lu}, \citenamefont {Zhang}, \citenamefont {Zhang},
  \citenamefont {Dong}, \citenamefont {Wang}, \citenamefont {Feng},\ and\
  \citenamefont {Zhang}}]{Chen2024PRXPinning}%
  \BibitemOpen
  \bibfield  {author} {\bibinfo {author} {\bibfnamefont {C.}~\bibnamefont
  {Chen}}, \bibinfo {author} {\bibfnamefont {Y.}~\bibnamefont {Liu}}, \bibinfo
  {author} {\bibfnamefont {Y.}~\bibnamefont {Chen}}, \bibinfo {author}
  {\bibfnamefont {Y.~N.}\ \bibnamefont {Hu}}, \bibinfo {author} {\bibfnamefont
  {T.~Z.}\ \bibnamefont {Zhang}}, \bibinfo {author} {\bibfnamefont
  {D.}~\bibnamefont {Li}}, \bibinfo {author} {\bibfnamefont {X.}~\bibnamefont
  {Wang}}, \bibinfo {author} {\bibfnamefont {C.~X.}\ \bibnamefont {Wang}},
  \bibinfo {author} {\bibfnamefont {Z.~Y.~W.}\ \bibnamefont {Lu}}, \bibinfo
  {author} {\bibfnamefont {Y.~H.}\ \bibnamefont {Zhang}}, \bibinfo {author}
  {\bibfnamefont {Q.~L.}\ \bibnamefont {Zhang}}, \bibinfo {author}
  {\bibfnamefont {X.~L.}\ \bibnamefont {Dong}}, \bibinfo {author}
  {\bibfnamefont {R.}~\bibnamefont {Wang}}, \bibinfo {author} {\bibfnamefont
  {D.~L.}\ \bibnamefont {Feng}},\ and\ \bibinfo {author} {\bibfnamefont
  {T.}~\bibnamefont {Zhang}},\ }\bibfield  {title} {\bibinfo {title} {Revealing
  the microscopic mechanism of elementary vortex pinning in superconductors},\
  }\href {https://doi.org/10.1103/PhysRevX.14.041039} {\bibfield  {journal}
  {\bibinfo  {journal} {Phys. Rev. X}\ }\textbf {\bibinfo {volume} {14}},\
  \bibinfo {pages} {041039} (\bibinfo {year} {2024})}\BibitemShut {NoStop}%
\bibitem [{\citenamefont {de~Gennes}(1966)}]{deGennes1966Superconductivity}%
  \BibitemOpen
  \bibfield  {author} {\bibinfo {author} {\bibfnamefont {P.~G.}\ \bibnamefont
  {de~Gennes}},\ }\href@noop {} {\emph {\bibinfo {title} {Superconductivity of
  Metals and Alloys}}}\ (\bibinfo  {publisher} {W. A. Benjamin},\ \bibinfo
  {address} {New York},\ \bibinfo {year} {1966})\BibitemShut {NoStop}%
\bibitem [{\citenamefont {Gygi}\ and\ \citenamefont
  {Schl{\"u}ter}(1991)}]{GygiSchluter1991VortexBdG}%
  \BibitemOpen
  \bibfield  {author} {\bibinfo {author} {\bibfnamefont {F.}~\bibnamefont
  {Gygi}}\ and\ \bibinfo {author} {\bibfnamefont {M.}~\bibnamefont
  {Schl{\"u}ter}},\ }\bibfield  {title} {\bibinfo {title} {Self-consistent
  electronic structure of a vortex line in a type-ii superconductor},\ }\href
  {https://doi.org/10.1103/PhysRevB.43.7609} {\bibfield  {journal} {\bibinfo
  {journal} {Phys. Rev. B}\ }\textbf {\bibinfo {volume} {43}},\ \bibinfo
  {pages} {7609} (\bibinfo {year} {1991})}\BibitemShut {NoStop}%
\bibitem [{\citenamefont {Kamihara}\ \emph {et~al.}(2008)\citenamefont
  {Kamihara}, \citenamefont {Watanabe}, \citenamefont {Hirano},\ and\
  \citenamefont {Hosono}}]{Kamihara2008IronBased}%
  \BibitemOpen
  \bibfield  {author} {\bibinfo {author} {\bibfnamefont {Y.}~\bibnamefont
  {Kamihara}}, \bibinfo {author} {\bibfnamefont {T.}~\bibnamefont {Watanabe}},
  \bibinfo {author} {\bibfnamefont {M.}~\bibnamefont {Hirano}},\ and\ \bibinfo
  {author} {\bibfnamefont {H.}~\bibnamefont {Hosono}},\ }\bibfield  {title}
  {\bibinfo {title} {Iron-based layered superconductor
  {La[O$_{1-x}$F$_x$]FeAs}~($x=0.05$--$0.12$) with {$T_c=26$}~{K}},\ }\href
  {https://doi.org/10.1021/ja800073m} {\bibfield  {journal} {\bibinfo
  {journal} {J. Am. Chem. Soc.}\ }\textbf {\bibinfo {volume} {130}},\ \bibinfo
  {pages} {3296} (\bibinfo {year} {2008})}\BibitemShut {NoStop}%
\bibitem [{\citenamefont {Huang}\ \emph {et~al.}(2017)\citenamefont {Huang},
  \citenamefont {Feng}, \citenamefont {Ni}, \citenamefont {Li}, \citenamefont
  {Hu}, \citenamefont {Liu}, \citenamefont {Mao}, \citenamefont {Zhou},
  \citenamefont {Zhou}, \citenamefont {Jin}, \citenamefont {Wang},
  \citenamefont {Yuan}, \citenamefont {Dong},\ and\ \citenamefont
  {Zhao}}]{Huang2017LiFeOHFeSeFilm}%
  \BibitemOpen
  \bibfield  {author} {\bibinfo {author} {\bibfnamefont {Y.}~\bibnamefont
  {Huang}}, \bibinfo {author} {\bibfnamefont {Z.}~\bibnamefont {Feng}},
  \bibinfo {author} {\bibfnamefont {S.}~\bibnamefont {Ni}}, \bibinfo {author}
  {\bibfnamefont {J.}~\bibnamefont {Li}}, \bibinfo {author} {\bibfnamefont
  {W.}~\bibnamefont {Hu}}, \bibinfo {author} {\bibfnamefont {S.}~\bibnamefont
  {Liu}}, \bibinfo {author} {\bibfnamefont {Y.}~\bibnamefont {Mao}}, \bibinfo
  {author} {\bibfnamefont {H.}~\bibnamefont {Zhou}}, \bibinfo {author}
  {\bibfnamefont {F.}~\bibnamefont {Zhou}}, \bibinfo {author} {\bibfnamefont
  {K.}~\bibnamefont {Jin}}, \bibinfo {author} {\bibfnamefont {H.}~\bibnamefont
  {Wang}}, \bibinfo {author} {\bibfnamefont {J.}~\bibnamefont {Yuan}}, \bibinfo
  {author} {\bibfnamefont {X.}~\bibnamefont {Dong}},\ and\ \bibinfo {author}
  {\bibfnamefont {Z.}~\bibnamefont {Zhao}},\ }\bibfield  {title} {\bibinfo
  {title} {Superconducting {(Li,Fe)OHFeSe} film of high quality and high
  critical parameters},\ }\href {https://doi.org/10.1088/0256-307X/34/7/077404}
  {\bibfield  {journal} {\bibinfo  {journal} {Chin. Phys. Lett.}\ }\textbf
  {\bibinfo {volume} {34}},\ \bibinfo {pages} {077404} (\bibinfo {year}
  {2017})}\BibitemShut {NoStop}%
\bibitem [{\citenamefont {Qin}\ \emph {et~al.}(2021)\citenamefont {Qin},
  \citenamefont {Chen}, \citenamefont {Guo}, \citenamefont {Pan}, \citenamefont
  {Zhang}, \citenamefont {Xu}, \citenamefont {Chen}, \citenamefont {He},
  \citenamefont {Mei}, \citenamefont {Chen}, \citenamefont {Ye},\ and\
  \citenamefont {Wang}}]{Qin2021FeTeBi2Te3}%
  \BibitemOpen
  \bibfield  {author} {\bibinfo {author} {\bibfnamefont {H.}~\bibnamefont
  {Qin}}, \bibinfo {author} {\bibfnamefont {X.}~\bibnamefont {Chen}}, \bibinfo
  {author} {\bibfnamefont {B.}~\bibnamefont {Guo}}, \bibinfo {author}
  {\bibfnamefont {T.}~\bibnamefont {Pan}}, \bibinfo {author} {\bibfnamefont
  {M.}~\bibnamefont {Zhang}}, \bibinfo {author} {\bibfnamefont
  {B.}~\bibnamefont {Xu}}, \bibinfo {author} {\bibfnamefont {J.}~\bibnamefont
  {Chen}}, \bibinfo {author} {\bibfnamefont {H.}~\bibnamefont {He}}, \bibinfo
  {author} {\bibfnamefont {J.}~\bibnamefont {Mei}}, \bibinfo {author}
  {\bibfnamefont {W.}~\bibnamefont {Chen}}, \bibinfo {author} {\bibfnamefont
  {F.}~\bibnamefont {Ye}},\ and\ \bibinfo {author} {\bibfnamefont
  {G.}~\bibnamefont {Wang}},\ }\bibfield  {title} {\bibinfo {title} {Moir{\'e}
  superlattice-induced superconductivity in one-unit-cell {FeTe}},\ }\href
  {https://doi.org/10.1021/acs.nanolett.0c04048} {\bibfield  {journal}
  {\bibinfo  {journal} {Nano Lett.}\ }\textbf {\bibinfo {volume} {21}},\
  \bibinfo {pages} {1327} (\bibinfo {year} {2021})}\BibitemShut {NoStop}%
\bibitem [{\citenamefont {Ren}\ \emph {et~al.}(2021)\citenamefont {Ren},
  \citenamefont {Ru}, \citenamefont {Peng}, \citenamefont {Li}, \citenamefont
  {Lu}, \citenamefont {Chen}, \citenamefont {Wang}, \citenamefont {Fang},
  \citenamefont {Li}, \citenamefont {Huang}, \citenamefont {Wang},
  \citenamefont {Wang},\ and\ \citenamefont {Li}}]{Ren2021OxygenFeTe}%
  \BibitemOpen
  \bibfield  {author} {\bibinfo {author} {\bibfnamefont {W.}~\bibnamefont
  {Ren}}, \bibinfo {author} {\bibfnamefont {H.}~\bibnamefont {Ru}}, \bibinfo
  {author} {\bibfnamefont {K.}~\bibnamefont {Peng}}, \bibinfo {author}
  {\bibfnamefont {H.}~\bibnamefont {Li}}, \bibinfo {author} {\bibfnamefont
  {S.}~\bibnamefont {Lu}}, \bibinfo {author} {\bibfnamefont {A.}~\bibnamefont
  {Chen}}, \bibinfo {author} {\bibfnamefont {P.}~\bibnamefont {Wang}}, \bibinfo
  {author} {\bibfnamefont {X.}~\bibnamefont {Fang}}, \bibinfo {author}
  {\bibfnamefont {Z.}~\bibnamefont {Li}}, \bibinfo {author} {\bibfnamefont
  {R.}~\bibnamefont {Huang}}, \bibinfo {author} {\bibfnamefont
  {L.}~\bibnamefont {Wang}}, \bibinfo {author} {\bibfnamefont {Y.}~\bibnamefont
  {Wang}},\ and\ \bibinfo {author} {\bibfnamefont {F.}~\bibnamefont {Li}},\
  }\bibfield  {title} {\bibinfo {title} {Oxygen adsorption induced
  superconductivity in ultrathin {FeTe} film on {SrTiO3}(001)},\ }\href
  {https://doi.org/10.3390/ma14164584} {\bibfield  {journal} {\bibinfo
  {journal} {Materials}\ }\textbf {\bibinfo {volume} {14}},\ \bibinfo {pages}
  {4584} (\bibinfo {year} {2021})}\BibitemShut {NoStop}%
\bibitem [{\citenamefont {Yan}\ \emph {et~al.}(2026)\citenamefont {Yan},
  \citenamefont {Wang}, \citenamefont {Xia}, \citenamefont {Paolini},
  \citenamefont {Chan}, \citenamefont {Dihingia}, \citenamefont {Rong},
  \citenamefont {Xiao}, \citenamefont {Halanayake}, \citenamefont {Song},
  \citenamefont {Gowda}, \citenamefont {Hickey}, \citenamefont {Wu},
  \citenamefont {Yu}, \citenamefont {Hirschfeld},\ and\ \citenamefont
  {Chang}}]{Yan2026StoichiometricFeTe}%
  \BibitemOpen
  \bibfield  {author} {\bibinfo {author} {\bibfnamefont {Z.-J.}\ \bibnamefont
  {Yan}}, \bibinfo {author} {\bibfnamefont {Z.}~\bibnamefont {Wang}}, \bibinfo
  {author} {\bibfnamefont {B.}~\bibnamefont {Xia}}, \bibinfo {author}
  {\bibfnamefont {S.}~\bibnamefont {Paolini}}, \bibinfo {author} {\bibfnamefont
  {Y.-T.}\ \bibnamefont {Chan}}, \bibinfo {author} {\bibfnamefont
  {N.}~\bibnamefont {Dihingia}}, \bibinfo {author} {\bibfnamefont
  {H.}~\bibnamefont {Rong}}, \bibinfo {author} {\bibfnamefont {P.}~\bibnamefont
  {Xiao}}, \bibinfo {author} {\bibfnamefont {K.~D.}\ \bibnamefont
  {Halanayake}}, \bibinfo {author} {\bibfnamefont {J.}~\bibnamefont {Song}},
  \bibinfo {author} {\bibfnamefont {V.}~\bibnamefont {Gowda}}, \bibinfo
  {author} {\bibfnamefont {D.~R.}\ \bibnamefont {Hickey}}, \bibinfo {author}
  {\bibfnamefont {W.}~\bibnamefont {Wu}}, \bibinfo {author} {\bibfnamefont
  {J.}~\bibnamefont {Yu}}, \bibinfo {author} {\bibfnamefont {P.~J.}\
  \bibnamefont {Hirschfeld}},\ and\ \bibinfo {author} {\bibfnamefont {C.-Z.}\
  \bibnamefont {Chang}},\ }\bibfield  {title} {\bibinfo {title} {Stoichiometric
  {FeTe} is a superconductor},\ }\href
  {https://doi.org/10.1038/s41586-026-10321-0} {\bibfield  {journal} {\bibinfo
  {journal} {Nature}\ }\textbf {\bibinfo {volume} {652}},\ \bibinfo {pages}
  {342} (\bibinfo {year} {2026})}\BibitemShut {NoStop}%
\bibitem [{\citenamefont {Gor'kov}(1959)}]{Gorkov1959GL}%
  \BibitemOpen
  \bibfield  {author} {\bibinfo {author} {\bibfnamefont {L.~P.}\ \bibnamefont
  {Gor'kov}},\ }\bibfield  {title} {\bibinfo {title} {Microscopic derivation of
  the {Ginzburg--Landau} equations in the theory of superconductivity},\ }\href
  {http://jetp.ras.ru/cgi-bin/e/index/e/9/6/p1364?a=list} {\bibfield  {journal}
  {\bibinfo  {journal} {Sov. Phys. JETP}\ }\textbf {\bibinfo {volume} {9}},\
  \bibinfo {pages} {1364} (\bibinfo {year} {1959})}\BibitemShut {NoStop}%
\bibitem [{\citenamefont {Bardeen}\ \emph {et~al.}(1957)\citenamefont
  {Bardeen}, \citenamefont {Cooper},\ and\ \citenamefont
  {Schrieffer}}]{Bardeen1957BCS}%
  \BibitemOpen
  \bibfield  {author} {\bibinfo {author} {\bibfnamefont {J.}~\bibnamefont
  {Bardeen}}, \bibinfo {author} {\bibfnamefont {L.~N.}\ \bibnamefont
  {Cooper}},\ and\ \bibinfo {author} {\bibfnamefont {J.~R.}\ \bibnamefont
  {Schrieffer}},\ }\bibfield  {title} {\bibinfo {title} {Theory of
  superconductivity},\ }\href {https://doi.org/10.1103/PhysRev.108.1175}
  {\bibfield  {journal} {\bibinfo  {journal} {Phys. Rev.}\ }\textbf {\bibinfo
  {volume} {108}},\ \bibinfo {pages} {1175} (\bibinfo {year}
  {1957})}\BibitemShut {NoStop}%
\bibitem [{\citenamefont {Shi}\ \emph {et~al.}(2026)\citenamefont {Shi},
  \citenamefont {Xie}, \citenamefont {Zhang}, \citenamefont {Chu},\ and\
  \citenamefont {Gong}}]{ShiPinningCode}%
  \BibitemOpen
  \bibfield  {author} {\bibinfo {author} {\bibfnamefont {H.}~\bibnamefont
  {Shi}}, \bibinfo {author} {\bibfnamefont {Y.}~\bibnamefont {Xie}}, \bibinfo
  {author} {\bibfnamefont {T.}~\bibnamefont {Zhang}}, \bibinfo {author}
  {\bibfnamefont {W.}~\bibnamefont {Chu}},\ and\ \bibinfo {author}
  {\bibfnamefont {X.-G.}\ \bibnamefont {Gong}},\ }\href
  {https://doi.org/10.5281/zenodo.21790252} {\bibinfo {title} {Pinning
  algorithm: First-principles quantum-spectral vortex pinning}} (\bibinfo
  {year} {2026})\BibitemShut {NoStop}%
\bibitem [{\citenamefont {Kohn}\ and\ \citenamefont
  {Sham}(1965)}]{KohnSham1965DFT}%
  \BibitemOpen
  \bibfield  {author} {\bibinfo {author} {\bibfnamefont {W.}~\bibnamefont
  {Kohn}}\ and\ \bibinfo {author} {\bibfnamefont {L.~J.}\ \bibnamefont
  {Sham}},\ }\bibfield  {title} {\bibinfo {title} {Self-consistent equations
  including exchange and correlation effects},\ }\href
  {https://doi.org/10.1103/PhysRev.140.A1133} {\bibfield  {journal} {\bibinfo
  {journal} {Phys. Rev.}\ }\textbf {\bibinfo {volume} {140}},\ \bibinfo {pages}
  {A1133} (\bibinfo {year} {1965})}\BibitemShut {NoStop}%
\bibitem [{\citenamefont {Perdew}\ \emph {et~al.}(1996)\citenamefont {Perdew},
  \citenamefont {Burke},\ and\ \citenamefont {Ernzerhof}}]{Perdew1996PBE}%
  \BibitemOpen
  \bibfield  {author} {\bibinfo {author} {\bibfnamefont {J.~P.}\ \bibnamefont
  {Perdew}}, \bibinfo {author} {\bibfnamefont {K.}~\bibnamefont {Burke}},\ and\
  \bibinfo {author} {\bibfnamefont {M.}~\bibnamefont {Ernzerhof}},\ }\bibfield
  {title} {\bibinfo {title} {Generalized gradient approximation made simple},\
  }\href {https://doi.org/10.1103/PhysRevLett.77.3865} {\bibfield  {journal}
  {\bibinfo  {journal} {Phys. Rev. Lett.}\ }\textbf {\bibinfo {volume} {77}},\
  \bibinfo {pages} {3865} (\bibinfo {year} {1996})}\BibitemShut {NoStop}%
\bibitem [{\citenamefont {Kresse}\ and\ \citenamefont
  {Furthm{\"u}ller}(1996)}]{Kresse1996VASP}%
  \BibitemOpen
  \bibfield  {author} {\bibinfo {author} {\bibfnamefont {G.}~\bibnamefont
  {Kresse}}\ and\ \bibinfo {author} {\bibfnamefont {J.}~\bibnamefont
  {Furthm{\"u}ller}},\ }\bibfield  {title} {\bibinfo {title} {Efficient
  iterative schemes for ab initio total-energy calculations using a plane-wave
  basis set},\ }\href {https://doi.org/10.1103/PhysRevB.54.11169} {\bibfield
  {journal} {\bibinfo  {journal} {Phys. Rev. B}\ }\textbf {\bibinfo {volume}
  {54}},\ \bibinfo {pages} {11169} (\bibinfo {year} {1996})}\BibitemShut
  {NoStop}%
\bibitem [{\citenamefont {Marzari}\ and\ \citenamefont
  {Vanderbilt}(1997)}]{MarzariVanderbilt1997Wannier}%
  \BibitemOpen
  \bibfield  {author} {\bibinfo {author} {\bibfnamefont {N.}~\bibnamefont
  {Marzari}}\ and\ \bibinfo {author} {\bibfnamefont {D.}~\bibnamefont
  {Vanderbilt}},\ }\bibfield  {title} {\bibinfo {title} {Maximally localized
  generalized {Wannier} functions for composite energy bands},\ }\href
  {https://doi.org/10.1103/PhysRevB.56.12847} {\bibfield  {journal} {\bibinfo
  {journal} {Phys. Rev. B}\ }\textbf {\bibinfo {volume} {56}},\ \bibinfo
  {pages} {12847} (\bibinfo {year} {1997})}\BibitemShut {NoStop}%
\bibitem [{\citenamefont {Mostofi}\ \emph {et~al.}(2008)\citenamefont
  {Mostofi}, \citenamefont {Yates}, \citenamefont {Lee}, \citenamefont {Souza},
  \citenamefont {Vanderbilt},\ and\ \citenamefont
  {Marzari}}]{Mostofi2008Wannier90}%
  \BibitemOpen
  \bibfield  {author} {\bibinfo {author} {\bibfnamefont {A.~A.}\ \bibnamefont
  {Mostofi}}, \bibinfo {author} {\bibfnamefont {J.~R.}\ \bibnamefont {Yates}},
  \bibinfo {author} {\bibfnamefont {Y.-S.}\ \bibnamefont {Lee}}, \bibinfo
  {author} {\bibfnamefont {I.}~\bibnamefont {Souza}}, \bibinfo {author}
  {\bibfnamefont {D.}~\bibnamefont {Vanderbilt}},\ and\ \bibinfo {author}
  {\bibfnamefont {N.}~\bibnamefont {Marzari}},\ }\bibfield  {title} {\bibinfo
  {title} {wannier90: A tool for obtaining maximally-localised {Wannier}
  functions},\ }\href {https://doi.org/10.1016/j.cpc.2007.11.016} {\bibfield
  {journal} {\bibinfo  {journal} {Comput. Phys. Commun.}\ }\textbf {\bibinfo
  {volume} {178}},\ \bibinfo {pages} {685} (\bibinfo {year}
  {2008})}\BibitemShut {NoStop}%
\bibitem [{\citenamefont {Marzari}\ \emph {et~al.}(2012)\citenamefont
  {Marzari}, \citenamefont {Mostofi}, \citenamefont {Yates}, \citenamefont
  {Souza},\ and\ \citenamefont {Vanderbilt}}]{Marzari2012WannierReview}%
  \BibitemOpen
  \bibfield  {author} {\bibinfo {author} {\bibfnamefont {N.}~\bibnamefont
  {Marzari}}, \bibinfo {author} {\bibfnamefont {A.~A.}\ \bibnamefont
  {Mostofi}}, \bibinfo {author} {\bibfnamefont {J.~R.}\ \bibnamefont {Yates}},
  \bibinfo {author} {\bibfnamefont {I.}~\bibnamefont {Souza}},\ and\ \bibinfo
  {author} {\bibfnamefont {D.}~\bibnamefont {Vanderbilt}},\ }\bibfield  {title}
  {\bibinfo {title} {Maximally localized {Wannier} functions: Theory and
  applications},\ }\href {https://doi.org/10.1103/RevModPhys.84.1419}
  {\bibfield  {journal} {\bibinfo  {journal} {Rev. Mod. Phys.}\ }\textbf
  {\bibinfo {volume} {84}},\ \bibinfo {pages} {1419} (\bibinfo {year}
  {2012})}\BibitemShut {NoStop}%
\bibitem [{\citenamefont {Kuhn}(1955)}]{Kuhn1955Hungarian}%
  \BibitemOpen
  \bibfield  {author} {\bibinfo {author} {\bibfnamefont {H.~W.}\ \bibnamefont
  {Kuhn}},\ }\bibfield  {title} {\bibinfo {title} {The {Hungarian} method for
  the assignment problem},\ }\href {https://doi.org/10.1002/nav.3800020109}
  {\bibfield  {journal} {\bibinfo  {journal} {Nav. Res. Logist. Q.}\ }\textbf
  {\bibinfo {volume} {2}},\ \bibinfo {pages} {83} (\bibinfo {year}
  {1955})}\BibitemShut {NoStop}%
\bibitem [{\citenamefont {Sch{\"o}nemann}(1966)}]{Schonemann1966Procrustes}%
  \BibitemOpen
  \bibfield  {author} {\bibinfo {author} {\bibfnamefont {P.~H.}\ \bibnamefont
  {Sch{\"o}nemann}},\ }\bibfield  {title} {\bibinfo {title} {A generalized
  solution of the orthogonal {Procrustes} problem},\ }\href
  {https://doi.org/10.1007/BF02289451} {\bibfield  {journal} {\bibinfo
  {journal} {Psychometrika}\ }\textbf {\bibinfo {volume} {31}},\ \bibinfo
  {pages} {1} (\bibinfo {year} {1966})}\BibitemShut {NoStop}%
\bibitem [{\citenamefont {Bl{\"o}chl}(1994)}]{Blochl1994PAW}%
  \BibitemOpen
  \bibfield  {author} {\bibinfo {author} {\bibfnamefont {P.~E.}\ \bibnamefont
  {Bl{\"o}chl}},\ }\bibfield  {title} {\bibinfo {title} {Projector
  augmented-wave method},\ }\href {https://doi.org/10.1103/PhysRevB.50.17953}
  {\bibfield  {journal} {\bibinfo  {journal} {Phys. Rev. B}\ }\textbf {\bibinfo
  {volume} {50}},\ \bibinfo {pages} {17953} (\bibinfo {year}
  {1994})}\BibitemShut {NoStop}%
\bibitem [{\citenamefont {Kresse}\ and\ \citenamefont
  {Joubert}(1999)}]{KresseJoubert1999PAW}%
  \BibitemOpen
  \bibfield  {author} {\bibinfo {author} {\bibfnamefont {G.}~\bibnamefont
  {Kresse}}\ and\ \bibinfo {author} {\bibfnamefont {D.}~\bibnamefont
  {Joubert}},\ }\bibfield  {title} {\bibinfo {title} {From ultrasoft
  pseudopotentials to the projector augmented-wave method},\ }\href
  {https://doi.org/10.1103/PhysRevB.59.1758} {\bibfield  {journal} {\bibinfo
  {journal} {Phys. Rev. B}\ }\textbf {\bibinfo {volume} {59}},\ \bibinfo
  {pages} {1758} (\bibinfo {year} {1999})}\BibitemShut {NoStop}%
\bibitem [{\citenamefont {Clem}(1975)}]{Clem1975VortexCore}%
  \BibitemOpen
  \bibfield  {author} {\bibinfo {author} {\bibfnamefont {J.~R.}\ \bibnamefont
  {Clem}},\ }\bibfield  {title} {\bibinfo {title} {Simple model for the vortex
  core in a type {II} superconductor},\ }\href
  {https://doi.org/10.1007/BF00116134} {\bibfield  {journal} {\bibinfo
  {journal} {J. Low Temp. Phys.}\ }\textbf {\bibinfo {volume} {18}},\ \bibinfo
  {pages} {427} (\bibinfo {year} {1975})}\BibitemShut {NoStop}%
\bibitem [{\citenamefont {Oliveira}\ \emph {et~al.}(1988)\citenamefont
  {Oliveira}, \citenamefont {Gross},\ and\ \citenamefont
  {Kohn}}]{OliveiraGrossKohn1988SCDFT}%
  \BibitemOpen
  \bibfield  {author} {\bibinfo {author} {\bibfnamefont {L.~N.}\ \bibnamefont
  {Oliveira}}, \bibinfo {author} {\bibfnamefont {E.~K.~U.}\ \bibnamefont
  {Gross}},\ and\ \bibinfo {author} {\bibfnamefont {W.}~\bibnamefont {Kohn}},\
  }\bibfield  {title} {\bibinfo {title} {Density-functional theory for
  superconductors},\ }\href {https://doi.org/10.1103/PhysRevLett.60.2430}
  {\bibfield  {journal} {\bibinfo  {journal} {Phys. Rev. Lett.}\ }\textbf
  {\bibinfo {volume} {60}},\ \bibinfo {pages} {2430} (\bibinfo {year}
  {1988})}\BibitemShut {NoStop}%
\end{thebibliography}%

\end{document}